\documentclass{pasj00}

\def\commenta{$^*$}
\def\commentb{$^\dagger$}
\def\commentc{$^\ddagger$}

\DeclareAbbreviation\AAHam{Astron. Abh. Hamburg. Sternw.}
\DeclareAbbreviation\AARv{Astron. Astrophys. Rev.}
\DeclareAbbreviation\AAS{American Astron. Soc. Meeting Abstracts}
\DeclareAbbreviation\AcA{Acta Astron.}
\DeclareAbbreviation\actaa{Acta Astron.}
\DeclareAbbreviation\Afz{Astrofizika}
\DeclareAbbreviation\AGAb{Astronomische Gesellschaft Abstract Ser.}
\DeclareAbbreviation\an{Astron. Nachr.}
\DeclareAbbreviation\AnAp{Annales d'Astrophysique}
\DeclareAbbreviation\AnTok{Tokyo Astron. Obs. Annals, Sec. Ser.}
\DeclareAbbreviation\Ap{Astrophysics}
\DeclareAbbreviation\ARep{Astron. Rep.}
\DeclareAbbreviation\AstBu{Astrophys. Bull.}
\DeclareAbbreviation\ATel{Astron. Telegram}
\DeclareAbbreviation\ATsir{Astron. Tsirk.}
\DeclareAbbreviation\AcApS{Acta Astrophys. Sinica}
\DeclareAbbreviation\AstL{Astron. Lett.}
\DeclareAbbreviation\BaltA{Baltic Astron.}
\DeclareAbbreviation\BANS{Bull. of the Astron. Institutes of the Netherlands Suppl. Ser.}
\DeclareAbbreviation\BASI{Bull. Astron. Soc. India}
\DeclareAbbreviation\BeSN{Be Newslett.}
\DeclareAbbreviation\BHarO{Harvard Coll. Obs. Bull.}
\DeclareAbbreviation\CBET{Cent. Bur. Electron. Telegrams}
\DeclareAbbreviation\CEAB{Central European Astrophys. Bull.}
\DeclareAbbreviation\ChJAA{Chinese J. of Astron. and Astrophys.}
\DeclareAbbreviation\caa{Chinese J. of Astron. and Astrophys.}
\DeclareAbbreviation\CoAsi{Asiago Contr.}
\DeclareAbbreviation\CoSka{Contributions of the Astronomical Observatory Skalnat\'e Pleso}
\DeclareAbbreviation\GCN{GRB Coord. Netw. Circ.}
\DeclareAbbreviation\ErgAN{Erg. Astron. Nachr.}
\DeclareAbbreviation\ibvs{IBVS}
\DeclareAbbreviation\IEEEP{IEEE Proc.}
\DeclareAbbreviation\JAD{J. Astron. Data}
\DeclareAbbreviation\JApA{J. of Astrophys. and Astron.}
\DeclareAbbreviation\JAVSO{J. American Assoc. Variable Star Obs.}
\DeclareAbbreviation\JBAA{J. Br. Astron. Assoc.}
\DeclareAbbreviation\JPhCS{J. of Physics Conference Series}
\DeclareAbbreviation\JPSJ{J. Phys. Soc. Japan}
\DeclareAbbreviation\JSARA{J. of the Southeastern Assoc. for Research in Astron.}
\DeclareAbbreviation\LowOB{Lowell Obs. Bull.}
\DeclareAbbreviation\MitAG{Mitteil. der Astronom. Gesell. Hamburg}
\DeclareAbbreviation\MitVS{Mitteil. Ver\"{a}nderl. Sterne}
\DeclareAbbreviation\MmSAI{Mem. Soc. Astron. Ital.}
\DeclareAbbreviation\memsai{Mem. Soc. Astron. Ital.}
\DeclareAbbreviation\Msngr{Messenger}
\DeclareAbbreviation\NewA{New Astron.}
\DeclareAbbreviation\na{New Astron.}
\DeclareAbbreviation\NewAR{New Astron. Rev.}
\DeclareAbbreviation\nar{New Astron. Rev.}
\DeclareAbbreviation\NInfo{Nauchnye Informatsii}
\DeclareAbbreviation\NPhS{Nature Physical Science}
\DeclareAbbreviation\OAP{Odessa Astron. Publ.}
\DeclareAbbreviation\Obs{Observatory}
\DeclareAbbreviation\OEJV{Open Eur. J. on Variable Stars}
\DeclareAbbreviation\PASA{Publ. Astron. Soc. Australia}
\DeclareAbbreviation\PASAu{Publ. Astron. Soc. Australia}
\DeclareAbbreviation\PAZh{Pis'ma AZh}
\DeclareAbbreviation\PJAB{Proc. Japan Acad. Ser. B}
\DeclareAbbreviation\POBeo{Publ. de l'Observatoire Astronomique de Beograd}
\DeclareAbbreviation\PCCP{Phys. Chem. Chem. Phys.}
\DeclareAbbreviation\PhR{Phys. Rep.}
\DeclareAbbreviation\PVSS{Publ. Variable Stars Sect. R. Astron. Soc. New Zealand}
\DeclareAbbreviation\PZ{Perem. Zvezdy}
\DeclareAbbreviation\PZP{Perem. Zvezdy, Prilozh.}
\DeclareAbbreviation\QJRAS{QJRAS}
\DeclareAbbreviation\RA{Ricerche Astronomiche}
\DeclareAbbreviation\RMxAA{Rev. Mexicana Astron. Astrof.}
\DeclareAbbreviation\RvMA{Reviews of Modern Astron.}
\DeclareAbbreviation\RvMP{Reviews of Modern Phys.}
\DeclareAbbreviation\SASS{Society for Astronom. Sciences Ann. Symp.}
\DeclareAbbreviation\Sci{Science}
\DeclareAbbreviation\SPIE{SPIE Proc.}
\DeclareAbbreviation\SvA{Soviet Astronomy}
\DeclareAbbreviation\SvAL{Soviet Astronomy Letters}
\DeclareAbbreviation\VeSon{Ver\"{o}ff. Sternw. Sonneberg}
\DeclareAbbreviation\VSOLJBul{VSOLJ Variable Star Bull.}
\DeclareAbbreviation\yCat{VizieR Online Data Catalog}
\DeclareAbbreviation\ZA{Z. Astrophys.}

\def\PublisherCambridge{Cambridge: Cambridge University Press}

\def\PublisherSpringer{Berlin: Springer}

\def\PublisherWorldScientific{Singapore: World Scientific Publishing}

\SetRunningHead{Y. Osaki and T. Kato}{Superoutbursts and Superhumps 
in SU UMa Stars}

\Received{2013/03/06}
\Accepted{2013/05/23}

\title{Study of Superoutbursts and  Superhumps in SU UMa Stars 
 by the Kepler Light Curves of V344 Lyrae and V1504 Cygni}

\author{Yoji \textsc{Osaki}}
\affil{Department of Astronomy, School of Science, University of Tokyo,
Hongo, Tokyo 113-0033}
\email{osaki@ruby.ocn.ne.jp}

\and

\author{Taichi \textsc{Kato}}
\affil{Department of Astronomy, Kyoto University,
       Sakyo-ku, Kyoto 606-8502}
\email{tkato@kusastro.kyoto-u.ac.jp}


\begin{document}

\maketitle

\begin{abstract}
We have studied the short-cadence Kepler public light curves of SU UMa stars, 
V344 Lyr and V1504 Cyg extending over a period of more than  
two years by using power spectral analysis. 
We determined the orbital period of V344 Lyr 
to be $P_{\rm orb}=0.087903(1)$~d. 
We also reanalyzed the frequency variation of the negative superhump 
in a complete supercycle of V1504 Cyg with additional data of 
the $O-C$ diagram, confirming that its characteristic variation is 
in accordance with the thermal-tidal instability model. 
We present a new two-dimensional period analysis based on 
a new method of a least absolute shrinkage and selection operator (Lasso).
The new method gives very 
sharp peaks in the power spectra, and it is very useful for studing of the 
frequency variation in cataclysmic variable stars. We also analyzed 
simultaneous frequency variations of the positive and negative superhumps. 
If they are appropriately converted, it is found that they vary in unison, 
indicating that they represent a disk-radius variation. 
We have also studied the frequency 
(or period) variations of positive superhumps during superoutbursts. 
These variations can be understood in a qualitative way by combining of 
the disk radius variation and the variation of pressure effects during a 
superoutburst. A sudden excitation of oscillation with a frequency range 
near to 
the negative superhump (which we call ``impulsive negative superhump'') 
was observed in the descending branch of several outbursts of V344 Lyr. 
These events seem to have occurred just prior to the next superoutburst,   
and to act as a ``lead'' of the impending superoutburst.  
\end{abstract}

\section{Introduction}

Dwarf novae are eruptive variable stars, showing quasi-periodic outbursts 
with a typical amplitude of 2--5 mag and a typical recurrence time of weeks 
to months. They belong to a more general class of variable stars called 
the cataclysmic variable stars, which are semidetached close binary systems 
consisting of a white-dwarf primary star and a red-dwarf secondary star. 
The mass transferred from the Roche-lobe filling secondary star 
does not accrete directly to the white dwarf, but it first forms an accretion 
disk around the primary white dwarf. The outburst of dwarf novae is now 
thought to be caused by a sudden brightening of the accretion disk.  
The outburst of ordinary dwarf novae is attributed to thermal instability 
in accretion disks 
(e.g., see \cite{can93DIreview}, \cite{las01DIDNXT} for review). 

An SU UMa stars is one subclass of dwarf novae with a short orbital 
period, which exhibits a long outburst called ``superoutburst'' 
having a brighter maximum by about 1 mag with a typical 
duration of about 2 weeks beside a short normal outburst with a duration of 
a few days of ordinary dwarf novae. The most enigmatic feature of 
SU UMa stars is an appearance of so-called superhump during 
the superoutburst, a periodic 
photometric hump with a period longer than the orbital period by a few 
percent. The superhump mentioned above is also called the positive superhump 
because some of SU UMa stars and nova-like variable stars exhibit periodic 
photometric hump with a period shorter than the orbital period 
by a few percent and the latter is called the negative superhump. 
General reviews of dwarf novae and SU UMa stars in particular are 
found in monographs dealing with the cataclysmic variable stars by 
\citet{war95book}, \citet{hel01book}.

The superoutbursts and superhumps in SU UMa stars are one of most intriguing 
phenomena in variable stars. The positive superhump is now thought to be 
produced by the eccentric precessing disk in which an accretion disk is 
deformed into eccentric form and its apsidal line precesses prograde and 
the positive superhump periodicity is produced by the synodic period between 
the progradely precessing disk and the orbiting secondary star. 
The (positive) superhump phenomenon is now well understood 
by the tidal instability in accretion disks; superhumps are produced by the 
periodic tidal stressing of the eccentric precessing accretion disk, which is 
in turn produced by the tidal 3:1 resonance instability between accretion disk 
flow and the orbiting secondary stars
(\cite{whi88tidal}, \cite{hir90SHexcess}, \cite{lub91SHa}). 
On the other hand, the negative superhump is thought to be produced 
by the tilted disk in which an accretion disk is tilted from the binary 
orbital plane and its nodal line precesses retrograde and the negative 
superhump periodicity is produced by the synodic period between the retrograde 
precessing tilted disk and the orbiting secondary star. 

As for the superoutburst phenomenon, \citet{osa89suuma} has proposed 
so-called thermal-tidal instability model (abbreviated as TTI model) 
in which the ordinary thermal instability is coupled with the tidal 
instability [see, \citet{osa96review} for a review of the TTI model].
Although the TTI 
model is fairly well accepted as a correct explanation of the superoutburst 
and supercycle phenomenon, objections to this model have been raised from 
time to time, and two alternative models have been proposed; an enhanced 
mass-transfer model (abbreviated EMT model) by \citet{sma91suumamodel}
and a pure thermal instability model by Cannizzo 
(see, \cite{can10v344lyr}).  Since we have 
already discussed about theoretical models for superoutbursts 
and superhumps in section 2 of \citet{osa13v1504cygKepler} (Paper I),
we do not repeat them here. 

The NASA Kepler mission (\cite{bor10Keplerfirst}; \cite{Kepler}) 
to search for the earth-like planets around stars has been a great success;  
besides its main objective it provides an unprecedented opportunity 
to study variable stars in great details  to the variable star community
and cataclysmic variable stars are no exception. Two SU UMa stars in the 
Kepler field of view have been observed with the short-cadence mode; they 
are V344 Lyr and V1504 Cyg.  In paper I, we have 
studied the Kepler public light curve of one of SU UMa stars, V1504 Cyg, 
and demonstrated that the Kepler light curve has revealed clear evidence 
that the superoutburst in this star is caused by the tidal instability and 
the superoutburst and supercycle phenomena in SU UMa stars can be basically 
explained by the thermal-tidal instability model. In this paper we extend 
our previous work and study another Kepler SU UMa star, V344 Lyr, 
together with 
some additional analysis of V1504 Cyg for data recently available to public.  

V344 Lyr is a dwarf nova in the Kepler field and its Kepler light curve was 
first examined by \citet{sti10v344lyr} who showed that the superhumps found 
during the superoutburst continue to be detected during the following 
quiescent state and the next normal outburst. By using the power spectral 
analysis, \citet{woo11v344lyr} have studied 
the positive and negative superhumps 
of this star by Kepler light curve of three quarters (June 2009 to 
March 2010). They have demonstrated that the ordinary superhump in the 
early phase of superoutburst is generated by viscous dissipation within 
the periodically flexing (eccentric) disk while in the later phase of 
the superoutburst the superhump signal 
(the so-called late superhump)\footnote{
   The origin of superhumps following the termination of
   the superoutburst is still in dispute (cf. \cite{Pdot}; 
   also subsection 4.2 in \cite{woo11v344lyr}).
   We refer to ``traditional late superhumps'' here, i.e. superhumps
   whose phase is $\sim$0.5 different from the superhumps during
   superoutbursts (see \cite{Pdot3} for the nomenclature.
}
is generated as the gas stream hot spot sweeps around the rim of the 
non-axisymmetric (eccentric) disk.

\section{The Kepler Light Curves of V344 Lyr and V1504 Cyg} 

In this paper we examine the short cadence Kepler public data of 
V344 Lyr and also V1504 Cyg, extending for two years and a quarter 
from June 2009 to September 2011 for about 830 d from Barycentric 
Jurian Date (BJD) 2455002 to 2455833. 
We use data of the Simple Aperture Photometry (SAP) whose count rates are 
converted to a relative magnitude, by ${\rm mag}=13-2.5\log e$ 
where $e$ is the Kepler electron count rate (electrons s$^{-1}$) 
and a constant of 13 is arbitrarily chosen just for convenience. 
The Kepler light curves of these two stars for about a period of 736~d 
have already been studied by \citet{can12v344lyr} and we also note 
that our data for V344 Lyr in the early part used here overlap with 
those of \citet{woo11v344lyr}. 

\begin{table*}
\caption{Superoutbursts and supercycles of V344 Lyr.\commenta}
        \label{tab:supercycleV344}
\begin{center}
\begin{tabular}{cccccccccc}
\hline
(1) SC & (2) start & (3) start & (4) end & (5) SC length  & (6) SO   & (7) SC & (8) number & (9) negative & (10) orbital \\
number & of SC\commentb & of SO\commentb & of SO\commentb & excluding SO\commentc & duration\commentc & length\commentc & of NO      &   SH         & signal \\
\hline
1 & -- & 55.5 & 73 & --& 17.5 & $>$70 & $>$4 & well visible  & no \\
2 & 73 & 161 & 178 & 88 & 17   & 105 &  7  & well visible & no \\
3 & 178  & 277   & 294  &  99  & 17 & 116   &  7  & no & visible \\
4 & 294  & 397 & 415  &  103 & 18 & 121   &   10  & no & visible \\
5 & 415 & 527   & 544.5 & 112   & 17.5   & 129.5 &  13  & no & visible  \\
6 & 544.5  & 641.5  & 659.5 & 97 &  18  &  115  & 10 & no & well visible \\
7 & 659.5 & 743   & 760  & 83.5 & 17 & 100.5 &   8  & well visible & no \\
8 & 760  & --  &  --  &  --   &  --  &   --  &  $>$9 &partly visible & partly 
visible  \\
\hline
  \multicolumn{10}{l}{\commenta Abbreviations in this table: supercycle (SC), superoutburst (SO), normal outburst (NO), superhump (SH).} \\
  \multicolumn{10}{l}{\commentb BJD$-$2455000.} \\
  \multicolumn{10}{l}{\commentc Unit: d.} \\
\end{tabular}
\end{center}
\end{table*}

\begin{table*}
\caption{Superoutbursts and supercycles of V1504 Cyg. \commenta}
        \label{tab:supercycleV1504}
\begin{center}
\begin{tabular}{cccccccccc}
\hline
(1) SC & (2) start & (3) start & (4) end & (5) SC length  & (6) SO   & (7) SC & (8) number & (9) negative & (10) orbital \\
number & of SC\commentb & of SO\commentb & of SO\commentb & excluding SO\commentc & duration\commentc & length\commentc & of NO      &   SH         & signal \\
\hline
1 &  --  & 74.5  & 88.5 &  --   & 14   & $>$88 & $>$8 & no & no \\
2 & 88.5 & 201   & 215  & 112.5 & 14   & 126.5 &  10  & no & no \\
3 & 215  & 312   & 325  &  97   & 13   & 110   &  9  & later half & partly \\
4 & 325  & 406.5 & 419  &  81.5 & 12.5 &  94   &   6  & full & visible \\
5 & 419  & 516   & 530  &  97   & 14   & 111   &   5  & full & partly \\
6 & 530  & 639  & 650 & 109& 11  & 120  & 10  & early part & later part \\
7 & 650 & 750   & 763.5  & 100 & 13.5 & 113.5 &   12  & no & yes \\
8 & 763.5  & --  &  --  &  --   &  --  &   --  &  $>$7 & no & yes \\
\hline
  \multicolumn{10}{l}{\commenta Abbreviations in this table: supercycle (SC), superoutburst (SO), normal outburst (NO), superhump (SH).} \\
  \multicolumn{10}{l}{\commentb BJD$-$2455000.} \\
  \multicolumn{10}{l}{\commentc Unit: d.} \\
\end{tabular}
\end{center}
\end{table*}

\subsection{Global light curves and the power spectra of V344 Lyr 
and V1504 Cyg}

\begin{figure*}
  \begin{center}
    \FigureFile(160mm,120mm){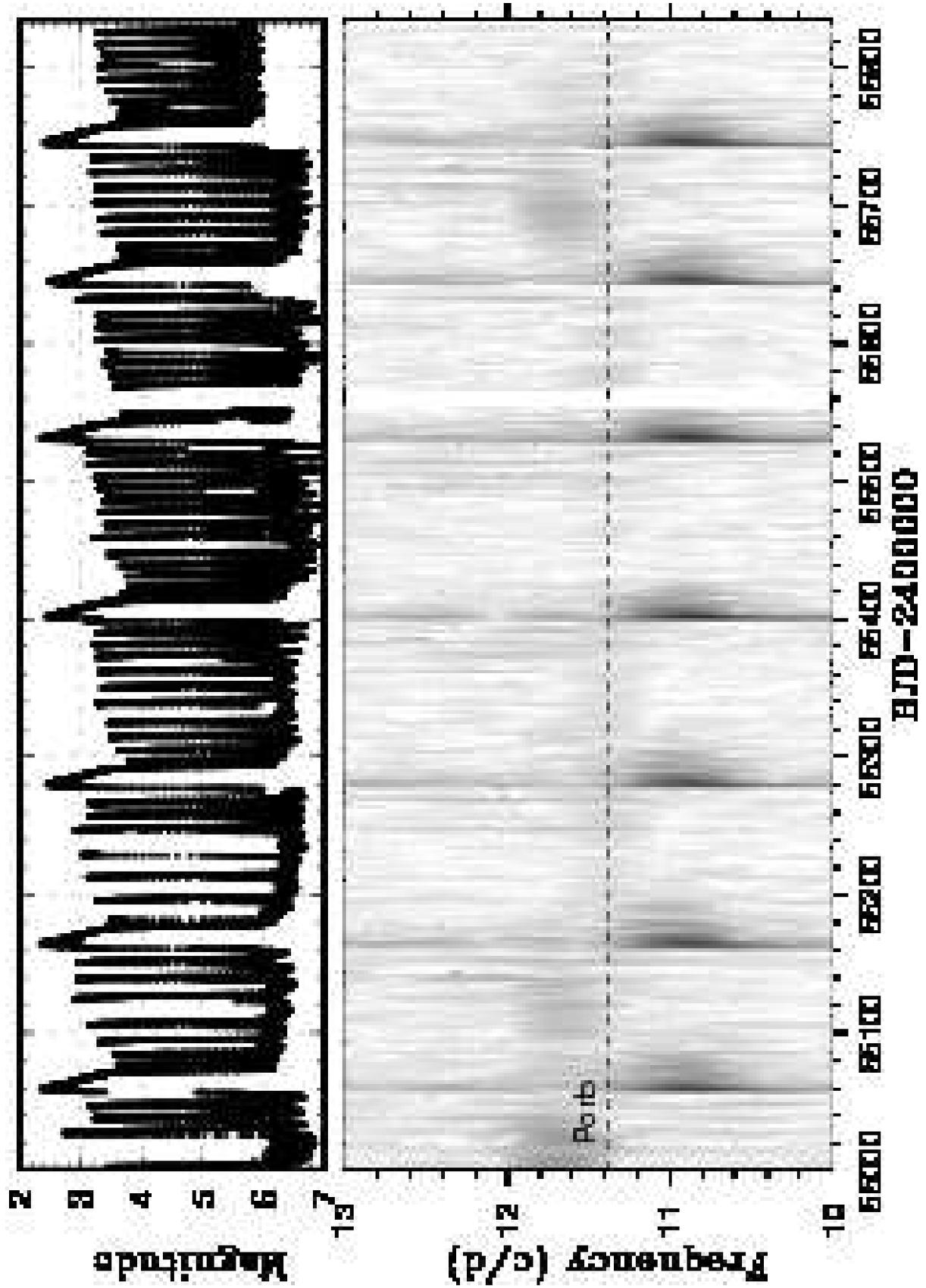}
  \end{center}
  \caption{Two-dimensional power spectrum of the Kepler light curve of 
  V344 Lyr for Q2-Q10.
  (upper:) light curve; the Kepler data were binned to 0.02~d,
  (lower:) power spectrum. The width of 
  the sliding window and the time step used are 5~d and 0.5~d,
  respectively.}
  \label{fig:v344spec2d}
\end{figure*}

\begin{figure*}
  \begin{center}
    \FigureFile(160mm,120mm){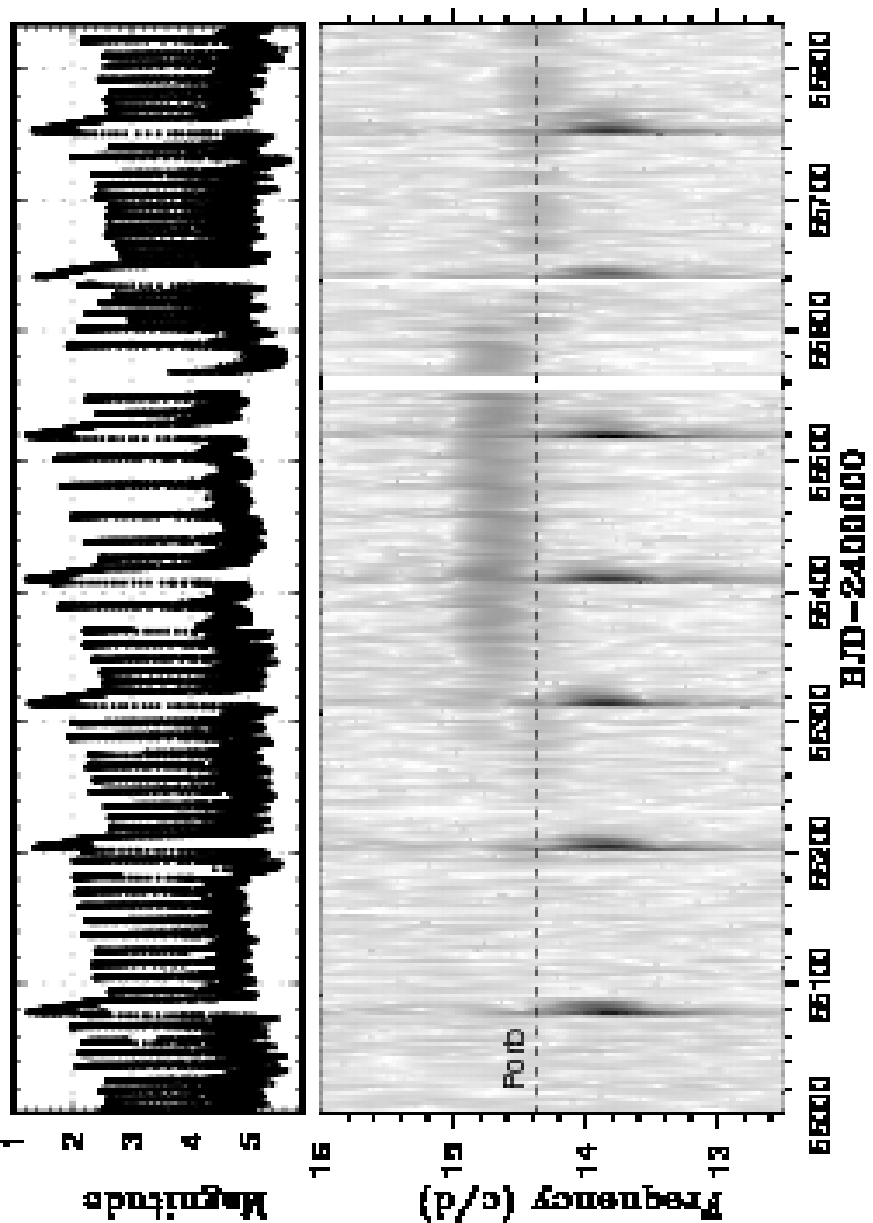}
  \end{center}
  \caption{Two-dimensional power spectrum of the Kepler light curve of 
  V1504 Cyg for Q2-Q10.
  (upper:) light curve; the Kepler data were binned to 0.02~d,
  (lower:) power spectrum. The width of 
  the sliding window and the time step used are 5~d and 0.5~d,
  respectively.}
  \label{fig:v1504spec2d}
\end{figure*}

We have first made two-dimensional power spectral analysis of the Kepler 
light curves of V344 Lyr and also of V1504 Cyg. 
In calculating the power spectra of V344 Lyr and 
V1504 Cyg, we have used the same method 
as that used in Paper I and its details should be consulted in Paper I.   
We show the overall power spectra together with its light curve for V344 Lyr 
in figure \ref{fig:v344spec2d} and for V1504 Cyg in figure
\ref{fig:v1504spec2d}, respectively. 

The Kepler data of V344 Lyr we have used here contain 7 superoutbursts and 
6 supercycles and we summarize their main characteristics 
in table \ref{tab:supercycleV344} as we did before for V1504 Cyg 
in Paper I. We also show them 
in table \ref{tab:supercycleV1504} for V1504 Cyg because of 
additional data for two quarters. 
The first column of tables 1 and 2 is the ordinal number while 
the next 3 columns are BJD date of the start of a supercycle (2), the start 
of the superoutburst (3), and its end (4), respectively, counted from 
BJD 2455000. The following 3 columns are lengths in days of a supercycle 
excluding superoutburst (5), of superoutburst duration (6), and of full 
supercycle (7). The next column (8) gives the number of normal outbursts 
within a supercycle while the last two columns (9) and (10) 
are comments on an appearance of negative superhumps and orbital signal 
in the power spectra. 

Let us now examine the light curve and power spectra of V344 Lyr 
in figure \ref{fig:v344spec2d} and table \ref{tab:supercycleV344}. 
Before we go into some details of the light curve, we first examine 
the orbital period of V344 Lyr. The orbital period of V344 Lyr was determined 
by \citet{woo11v344lyr} to be $P_{\rm orb}=0.087904$~d (2.11~hr or 11.38~c/d 
in frequency) from the Q4 data (our SC No. 3 with 200--275~d). 
Since our data contain the signal of 
the orbital humps more on the other occasions with SC No. 4 (320--370~d), 
SC No. 5 (460--520~d), and SC No. 6 (570--630~d), 
we re-determined the orbital period 
by the PDM method \citep{PDM} resulting in 0.0879049(11)~d 
for SC No. 3, 0.0879024(15)~d 
for SC No. 4, 0.0879014(36)~d for SC No. 5, 
and 0.0878937(15)~d for SC No. 6. 
Since the last value significantly differs from the first three, 
we determined from the first three that $P_{\rm orb}=0.087903(1)$~d, 
which is consistent with that of \citet{woo11v344lyr}. 
More recently, \citet{how13KeplerCVs} reported a short (3.4~hr)
radial-velocity study whose result is consistent with the photometric 
orbital period, although the inferred inclination (5--10$^\circ$)
appears to be too low to produce photometric orbital humps.

We note that the Kepler light curve of V344 Lyr and its power spectra 
show both similarity and difference with those of V1504 Cyg discussed 
in Paper I. The average supercycle length is about 114~d which is 
very close to that of V1504 Cyg (112~d). 
The strongest signal in the power spectra is that of positive superhumps 
with frequency around 10.8~c/d whenever a superoutburst occurs. Although the 
negative superhumps appear in the power spectra of V344 Lyr, these are more 
patchy (i.e., they come and go) than those of V1504 Cyg. We find them in the 
middle of SC No.1, SC No.2, and SC No.7. 

One of the most important findings about V1504 Cyg in Paper I was that 
a strong correlation between an appearance of negative superhumps and 
the length of quiescent interval between two consecutive normal outbursts 
in V1504 Cyg, that is, an appearance of well visible negative superhump 
tended to reduce 
the frequency of occurrence of normal outbursts. This leads us to classify 
supercycles of V1504 Cyg into two types of Type L and Type S with or 
without negative superhumps.  The same tendency is also seen in V344 Lyr 
for the SC No.1 and SC No.2 in figure \ref{fig:v344spec2d} when 
the negative superhump signal appears. 
In particular a very strong signal of negative superhumps appears in 
the first 40 days and it apparently suppresses occurrence of 
normal outburst (as already noted by \cite{can12v344lyr}). 

However the quiescence interval in SC No.3 is relatively long even though 
there is no visible signal of negative superhump. The frequency of normal 
outbursts in SC No.7 is not low even though the signal of negative superhumps 
is quite visible. Nevertheless the number of normal outbursts in SC No.7 
is 8 and it is relatively small because it is 10 for SC No.4, 13 for SC No.5, 
and 10 for SC No.6, where no signal of the negative superhump is visible 
in the power spectrum.
The correlation between appearance of negative superhump and 
quiescence interval is rather weak in the case of V344 Lyr. 
Since this correlation is fairly well established in V1504 Cyg as well 
as in ER UMa (\cite{ohs12eruma}, \cite{zem13eruma}) and
V503 Cyg (\cite{kat02v503cyg}, \cite{Pdot4}),
it is not clear whether V344 Lyr is an exceptional case or not. 

Let us now examine the light curve and power spectra of V1504 Cyg 
for additional data. The newly added two quarters of 
the Kepler data basically confirm the finding of Paper I as seen in 
figure \ref{fig:v1504spec2d} 
and table \ref{tab:supercycleV1504}, that is, the supercycle No. 7, 
which shows no signal of negative 
superhumps but instead a strong signal of the orbital hump, has the biggest 
number of normal outbursts of 12 in a supercycle, fitting with the 
classification of the Type S supercycle. Thus there is no need to revise 
our conclusion about V1504 Cyg with new data.

\subsection{Frequency Variation of the Negative Superhumps during Supercycles}

One of the most important findings in Paper I is that the frequency of 
the negative superhump (nSH) varies systematically during a supercycle 
in V1504 Cyg. As discussed in Paper I, if we adopt a tilted disk 
as the origin of 
the negative superhump, the frequency of negative superhump is given 
by a synodic frequency between a retrograde-precessing tilted disk 
and the orbiting secondary star, and it is written as  

\begin{equation}
\frac{\nu_{\rm nSH}}{\nu_{\rm orb}}
=1+\frac{3}{7} \frac{q}{\sqrt{1+q}} 
(\frac{R_d}{A})^{3/2} \cos \theta, 
\label{equ:fr_nSH}
\end{equation}
where $\nu_{\rm nSH}$, and $\nu_{\rm orb}$ are frequencies of negative 
superhump, and of the binary orbit, respectively,  $\theta$ is a tilt 
angle of the disk, $q=M_2/M_1$ is the binary mass ratio, and $A$ and $R_d$ 
are the binary separation and the disk radius, respectively. 
Here we assume $\cos \theta \simeq 1$ for a slightly tilted disk, 
as we did in Paper I.
Observed frequency variation shown in figure 5 of Paper I, if interpreted 
as an indicator of the disk radius, agreed very well with the variation 
in disk radius predicted by the thermal-tidal instability 
model (\cite{osa89suuma}, \cite{osa05DImodel}).

Recently \citet{sma13negSH} criticized our results on frequency variations 
of negative superhumps during normal outburst cycle. 
He stated that in our figure 5 the minima of $\nu_{\rm nSH}$ occurred 
$\sim 3$~d {\it before} the initial rise to outburst maximum and the 
following increase of $\nu_{\rm nSH}$ till its maximum lasted 
for $\sim 3$~d while the model calculations for dwarf nova outbursts show  
that expansion of the disk occurs nearly {\it simultaneously} with the rising 
light and lasts for only $\sim 0.5$~d (which corresponds to a viscous 
time scale).  

\begin{figure*}
  \begin{center}
    \FigureFile(160mm,100mm){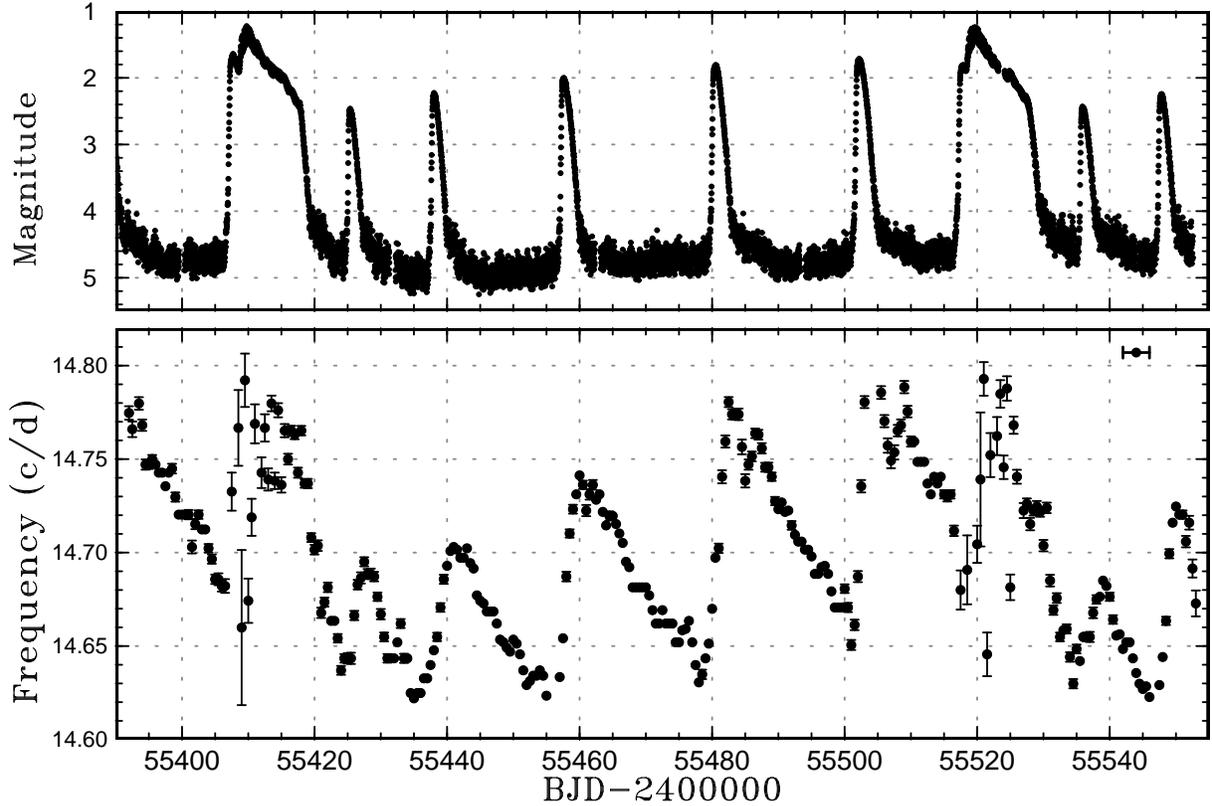}
  \end{center}
  \caption{Time evolution of frequency of the negative superhump covering 
  a complete supercycle No. 5 from BJD 2455390 to 2455550 of V1504 Cyg.
  The upper panel shows light curve while the lower panel does variation 
  in negative superhump frequency in units of cycle per day. The frequency 
  (or period) was calculated by using the PDM method 
  with a window width of 4~d and a time step of 0.5~d. The window width is 
  indicated as a horizontal bar at the upper right corner of the lower panel. 
  The Kepler data were averaged to 0.0005 d bins and the error bars
  in the lower panel represent 1-$\sigma$ errors in the periods.}
  \label{fig:v1504negshvar}
\end{figure*}

Since we think that this is a very important dispute, here we give some 
detailed account on this problem by reproducing figure 5 of paper I.
Figure \ref{fig:v1504negshvar} reproduces exactly figure 5 of Paper I. 
In calculating 
the local frequency $\nu$ (a frequency at a given time) in this figure, 
we used a method of sliding window with a window width 4~d 
and time step of 0.5~d.  
Now we examine Smak's criticism. 
First of all, we must apologize that we made a mistake of 2~d in time 
axis of our lower panel of figure 5 in the first version of astro-ph, 
arXiv:1212.1516v1 submitted on December 7, 2012, and the mistake 
was corrected in version 2 submitted on January 6, 2013 and 
figure \ref{fig:v1504negshvar} shown 
in this paper is the corrected one. This mistake crept in, when we 
inadvertently used time of local frequency at that of the starting date
of window instead of the correct one (i.e., the middle of the window).
We thank Dr. Smak for his careful scrutiny of our Paper I in its first 
version. By correcting this mistake, Smak's criticism reads that 
the minima of $\nu_{\rm nSH}$ occurred $\sim 1$~d {\it before} the initial 
rise to outburst and the following rise till its maximum lasted 
for $\sim 3$ days.

Let us now examine what happened in two normal outbursts at around day 458 
and around day 480 (the date is counted from BJD 2455000) in 
figure \ref{fig:v1504negshvar}
where a large jump in frequency occurred.  As can be clearly seen 
in figure \ref{fig:v1504negshvar}, the frequency jump from the 
local minima to the maxima in these two 
outbursts occurred with just 8 time-steps, 
that is, with 4~d, because our time-step is 0.5~d. 
This timescale of 4~d does not reflect the viscous tim scale of the disk 
but rather it simply reflects the window width of 
our calculations, an artifact of our calculations. That is, in determining 
a frequency at a given time, data with two days before and two days 
after the given time affected frequency determination of our calculation. 
This means that the frequency jump occurred in the middle of this jump and 
that the middle points in two outbursts exactly match their rising branches. 
The shape of this frequency jump in figure \ref{fig:v1504negshvar}
suggests that jumps occurred 
with a timescale less than the time-step (i.e., 0.5~d). 
In fact we recalculated the frequency variation 
by using a different window width 2~d and time step 0.5~d 
and we found in this case that the frequency 
jump occurred with 4 time-steps, i.e.,  2~d, 
which is again equal to the window width used. 
We conclude from figure \ref{fig:v1504negshvar} that a frequency jump 
occurred in the rising branch 
of normal outbursts and the timescale of this jump is less than 0.5~d; this is 
a conclusion which is in good agreement with the model calculations 
that Smak quoted. 

\begin{figure}
  \begin{center}
    \FigureFile(88mm,110mm){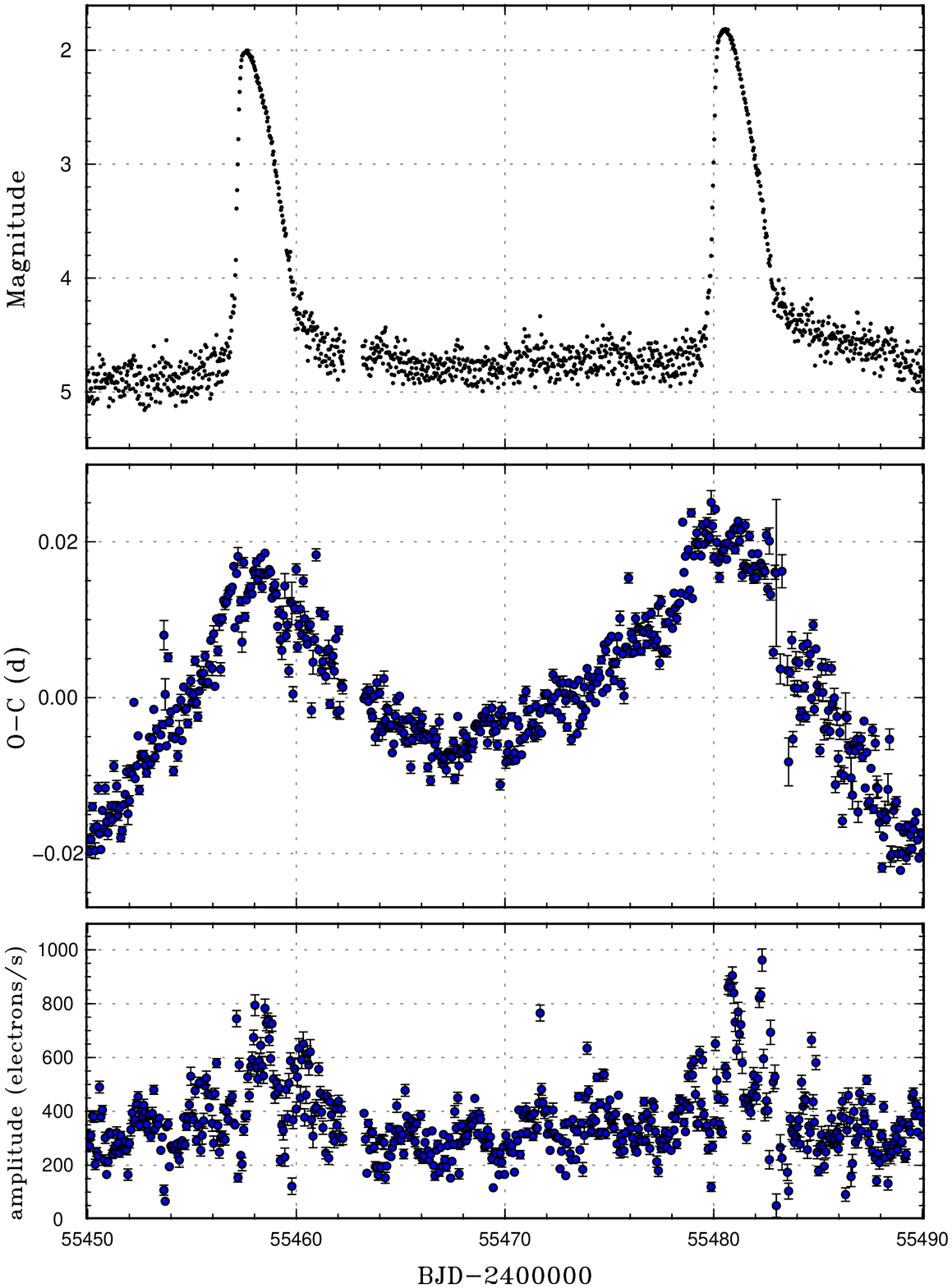}
  \end{center}
  \caption{ The $O-C$ diagram of negative superhumps and their 
  amplitudes for two outbursts of our interest in V1504 Cyg. 
  
  (Upper:) A light curve of V1504 Cyg for a period of
  BJD 2455450--2455490.  The Kepler data were averaged to 0.03~d bins.
  (Middle:) $O-C$ diagram of negative superhumps. A period of 0.0680626~d
  was used to draw this figure.
  (Lower:) Full amplitudes of negative superhumps in flux
  (electrons s$^{-1}$).
  The amplitudes were determined with the method in \citet{Pdot}
  using a sinusoid. The error bars in the middle and lower panels 
  represent 1-$\sigma$ errors.
  }
  \label{fig:v1504cygoc}
\end{figure}

To confirm this further, we calculated the $O-C$ diagram for 
the negative superhump of 
V1504 Cyg for a period BJD 2455450--2455490 including two normal 
outbursts, and figure \ref{fig:v1504cygoc}
exhibits its light curve, the $O-C$ diagram, 
and the amplitude of negative superhump from top to bottom, respectively.  
Following is the ephemeris of negative superhump maxima, from which 
$O-C$ is counted, 
\begin{equation}
{\rm BJD(max)}=2455449.9748+0.068067 E
\label{equ:Ephemeris_nSH}
\end{equation}

As seen from figure \ref{fig:v1504cygoc}, the $O-C$ diagram shows 
a cusp-like structure whenever 
an outburst occurs, indicating a jump in frequency of nSH. 
The peaks of cusp in $O-C$ diagram (and therefore jump in frequency) occurred 
at around BJD 2455457 and BJD 2455480, corresponding to the rising 
branch of each outburst and a jump in frequency occurs with a short 
time scale less than a day and most likely less than 0.5 d. 
This is completely consistent with the prediction of the disk 
instability model.

 The fact that a frequency jump occurs during a 
rising branch of an outburst strongly suggests that the outburst is 
of ``outside-in'', an unfavorable situation for Cannizzo's pure thermal 
instability model for the superoutburst of SU UMa stars \citep{can10v344lyr} 
because an ``inside-out'' outburst is required to explain a short outburst 
in their model.

\begin{figure}
  \begin{center}
    \FigureFile(88mm,140mm){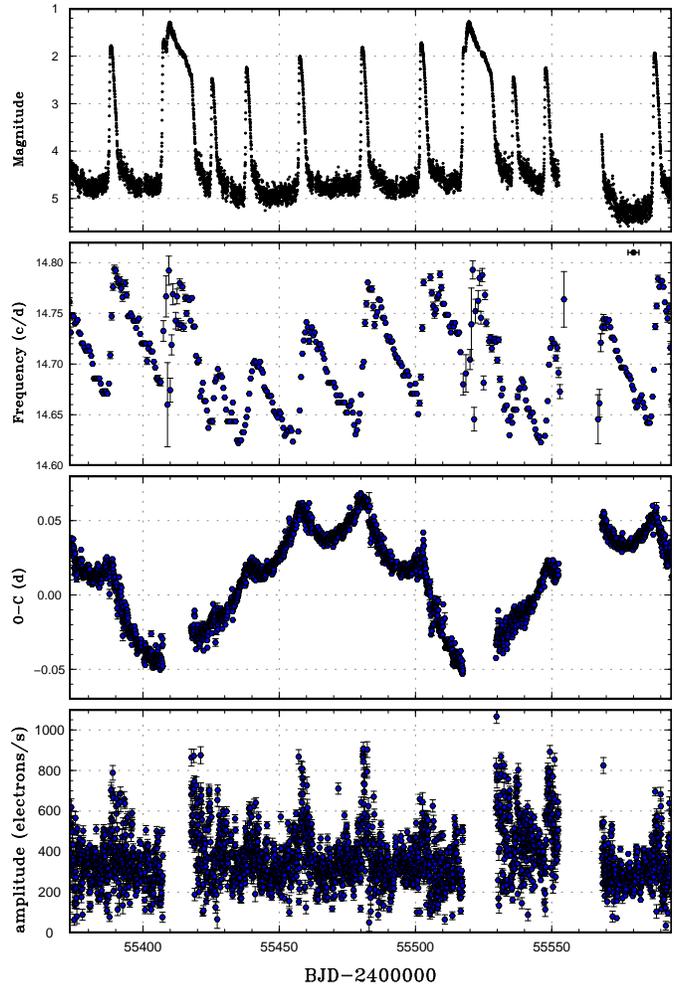}
  \end{center}
  \caption{The $O-C$ diagram of negative superhumps and their amplitude 
  variations in a complete supercycle No. 5 of V1504 Cyg.
  
  From top to bottom:
  (1:) A light curve of V1504 Cyg for a period of
  BJD 2455373--2455594.  The Kepler data were averaged to 0.03~d bins.
  (2:) Frequency of negative superhumps determined with PDM (4~d window).
  The window width is indicated by a horizontal bar at the upper right corner 
  and the error bars represent 1-$\sigma$ errors in the periods.
  (3:) $O-C$ diagram of negative superhumps.  A period of 0.0680626~d
  was used to draw this figure.
  (4:) Amplitudes of negative superhumps in flux (electrons s$^{-1}$).
  The error bars in the lowest two  panels represent 1-$\sigma$ errors.
  }
  \label{fig:v1504cygnshamp}
\end{figure}

To study further the $O-C$ diagram for a complete supercycle, 
we have calculated it 
together with amplitude variation for V1504 Cyg 
from BJD 2455373 to 2455594, including a supercycle No. 5 
which we studied in Paper I. Figure \ref{fig:v1504cygnshamp} 
illustrates the $O-C$ diagram for negative superhump 
with the same ephemeris of equation (\ref{equ:Ephemeris_nSH}) 
and its amplitude variation together with the light curve 
and its frequency variation. The $O-C$ diagram for a complete supercycle 
in figure \ref{fig:v1504cygnshamp} shows a characteristic variation. 
If we make a smooth curve (a spline curve) passing 
at each cusp point corresponding 
to each normal outburst, the resultant curve exhibits a convex form, implying 
$\dot P_{\rm nSH}<0$ in a long time scale during a supercycle, superimposed 
on it a short time-scale variation with a concave form 
between two consecutive normal outbursts, implying $\dot P_{\rm nSH}>0$; 
a phenomenon already discussed as a variation in frequency of nSH 
and seen in the second panel of figure \ref{fig:v1504cygnshamp}.
This indicates a secular trend of expansion in the disk radius in 
a supercycle, superimposed on it a decrease in the disk radius 
between two consecutive normal outbursts in a shorter time-scale; 
a phenomenon exactly predicted by the thermal-tidal instability model.

\subsection{Amplitude Variation of the Negative Superhump}

Let us now turn our attention to amplitude variation. 
 As seen in figure \ref{fig:v1504cygnshamp}, the amplitude of 
negative superhump exhibits a characteristic variation during supercycle 
No. 5. The amplitude variation of nSH is, however, very complex 
because quite different phenomena affect on amplitudes of nSH. 
In the following discussions we adopt the standard interpretation 
of the negative superhump, that is, the negative superhump is produced 
by a tilted disk which precesses retrograde by the secondary tidal torque.

The light source of the negative superhump {\it in quiescence} 
is thought to be a release of kinetic energy of gas stream 
by hitting different radius of the disk as gas stream sweeps 
around a tilted disk \citep{woo11v344lyr}.
First of all, we take note that we still do not understand 
the basic mechanism that causes 
and maintains disk tilt although several different mechanisms have 
been suggested about the origin of disk tilt, 
such as the 3:1 resonance \citep{lub92tilt}, 
magnetic coupling between the disk and either the secondary 
star or the primary white dwarf \citep{mur02warpeddisk}, stream-disk 
interaction with variable vertical component of stream 
due to asymmetric irradiation of the secondary star 
\citep{sma09negativeSH}, and 
dynamical lift by the gas stream \citep{mon10disktilt}.
As seen in global power spectra in figure \ref{fig:v344spec2d}
for V344 Lyr and in figure \ref{fig:v1504spec2d} 
for V1504 Cyg, the negative superhumps come and go in a long time-scale 
(in the case of V1504 Cyg a time scale more than 300~d, 
much longer than a supercycle). 
This long time-scale variation in amplitude of nSH is mostly likely 
produced by variation in tilt angle $\theta$ \citep{mon09negativeSH},  
in other words, variation in amplitude of a tilt mode of the disk.  

\begin{figure}
  \begin{center}
    \FigureFile(88mm,110mm){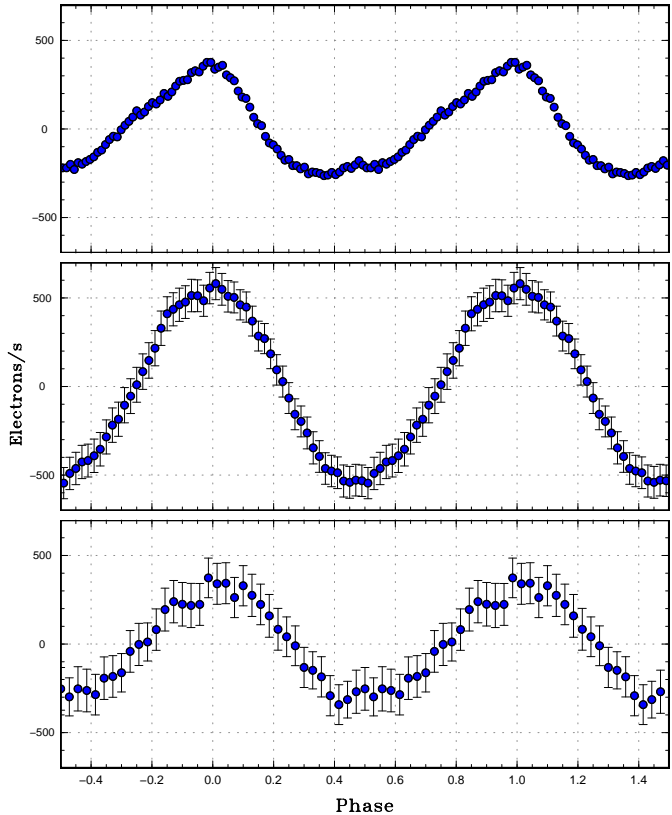}
  \end{center}
  \caption{Comparison of profiles of negative superhumps in V1504 Cyg,
  The light curves are given in unit electrons s$^{-1}$.
  (Upper:) Quiescence (BJD 2450440--2450448).
  (Middle:) Superoutburst (BJD 2450409--2450418).
  (Lower:) Normal outburst (BJD 2450387.5--2450390).
  The error bars represent 1-$\sigma$ errors. 
  The phases in these figures are chosen in such a way that the maximum light 
  occurs at phase zero for each panel. 
  }
  \label{fig:v1504profcomp}
\end{figure}

Let us now examine amplitude variation of nSH during a supercycle No. 5 
of V1504 Cyg in figure \ref{fig:v1504cygnshamp}. 
For a moment we assume that the tilt angle will 
stay more or less constant during this supercycle, 
although we must keep always in mind that the tilt angle can vary within a 
supercycle. From figure 5 we find that its amplitude increases 
whenever an outburst occurs. In particular, we note that 
profile of amplitude variation during a normal outburst mimics outburst 
profile of the light curve on a smaller scale. Although it is 
generally thought that the light source of negative superhump is due to a 
release of kinetic energy of the gas stream, these observations suggest 
that the disk component may contribute to the light variation 
of negative superhump during an outburst besides the gas stream component. 
A possible origin of the disk component to the negative superhump 
light source during the superoutburst has already been discussed 
in the Appendix of Paper I. 
We also recall that \citet{woo09negativeSH} noted that in their 
hydro-simulation of a tilted disk a negative 
superhump light signal was seen  even when the gas stream 
has been completely shut off. It is well known that the main light source 
of the ordinary (positive) superhumps is due to the viscous dissipation 
of tidally stressed eccentric disk (or the periodically 
flexing disk in the words of \cite{woo11v344lyr}) 
in the main part of the superoutburst while it is due to 
the gas stream in the late stage of the superoutburst and in quiescence as it 
is called the late superhump. 

We thus suggest that the amplitude increase of nSH during outbursts shown 
in figure \ref{fig:v1504cygnshamp} may be most likely produced by 
an addition of the disk component besides that of gas stream.
To confirm this, we compare phase averaged light curves of nSH 
in quiescence, at superoutburst No. 4, and at a normal outburst which occurred 
around 389 in figure \ref{fig:v1504profcomp}. 
As discussed already in paper I and shown in figure 7 and figure 8 of Paper I, 
the light curve of nSH in quiescence takes a saw-tooth like form 
with a rise time roughly twice the fall time in quiescence while it is more or 
less of sinusoidal form during a superoutburst. As seen in 
figure \ref{fig:v1504profcomp}, 
the light curve in a normal outburst takes more or less a form just between  
these two. If we accept our assumption about  
an addition of disk component to nSH light source besides the gas stream 
component in outbursts, amplitude variation of nSH shown in figure 5 is 
most naturally explained as it faithfully follows variation of light curve.

In his recent paper, \citet{sma13negSH} has argued that our results 
on the disk radius variation were inconsistent with observed amplitudes of 
negative superhumps in superoutburst and in quiescence. He assumed 
that light source of nSH was solely due to gas stream both 
in a superoutburst and in quiescence. We think his criticism is irrelevant 
because light source of nSH is most likely different between these two phases. 

As seen in figure \ref{fig:v1504cygnshamp}, the amplitude of nSH is 
suppressed when a strong positive superhump signal appears 
at the start of superoutburst. 
An excitation of the positive superhump (pSH) seems 
somehow to suppress temporarily 
the amplitude of nSH, although the exact cause of suppression of negative 
superhumps is not known. We do not know yet whether it is due to some 
physical mechanism or it is simply due to  an artifact of data analysis. 
Here we leave it as a problem to be solved in future.

\subsection {Two-dimensional Period Analysis using a New Period Analysis 
Called ``Lasso''} 

In paper I and figures \ref{fig:v344spec2d} and \ref{fig:v1504spec2d} 
above, we have used two-dimensional discrete 
Fourier transform (2D DFT) method in order to make power spectra. 
Here we introduce another method of two-dimensional spectral analysis 
by using a new method of period analysis called 
least absolute shrinkage and selection operator (``lasso'', \cite{lasso})
introduced to analysis of astronomical time-series data
\citep{kat12perlasso} which is very suitable to find peaks in power spectra 
(see also, \cite{kat13j1924}).
The advantage of the lasso analysis is that  peaks in power spectra 
are very sharp and thus it is very powerful in analyzing
rapidly changing periods as in outbursting dwarf novae.
 There is, however, a set-back in this method in that 
the resultant powers are not linear in power amplitude.
Thus this method and the Fourier transform method are complementary  
with each other. 
  
\begin{figure*}
  \begin{center}
    \FigureFile(160mm,120mm){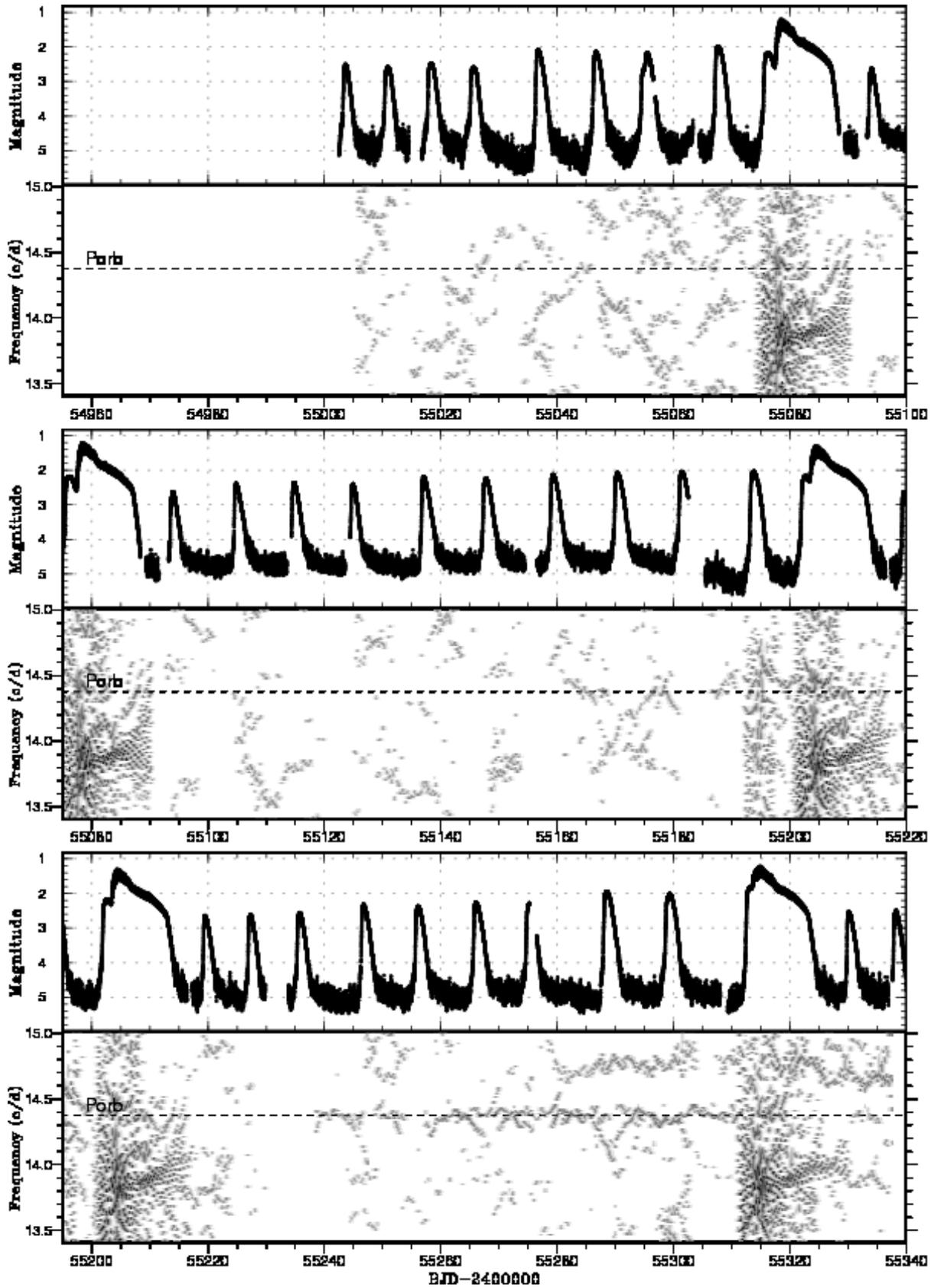}
  \end{center}
  \caption{Two-dimensional lasso power spectrum of the Kepler
  light curve of V1504 Cyg.
  (upper:) light curve; the Kepler data were binned to 0.02~d,
  (lower:) lasso power spectrum ($\log \lambda=-3.9$). The width of 
  the sliding window and the time step used are 5~d and 0.5~d,
  respectively.}
  \label{fig:v1504spec2dlasso}
\end{figure*}

In figure \ref{fig:v1504spec2dlasso} we show the two-dimensional 
power spectra of individual supercycles for V1504 Cyg by using 
this method and in figure \ref{fig:v344spec2dlasso} we do  
the same ones for V344 Lyr. Since the results for the first six supercycles 
of V1504 Cyg shown in figure \ref{fig:v1504spec2dlasso} 
overlap with those obtained by the discrete 
Fourier transform method given in figure 3 of Paper I, we can compare 
the results of these two methods.
Since much is common between figures \ref{fig:v1504spec2dlasso} and
\ref{fig:v344spec2dlasso}, we only discuss figure
\ref{fig:v1504spec2dlasso} here.  In the Lasso spectrum, nSH
and orbital modulation are always seen as clear separate
signals when nSH is present (e.g., BJD 2455310--2455420),
while the Fourier spectrum can barely separate them.
The Lasso spectrum clearly shows that the frequency of
the nSH rises quickly when normal outbursts start,
and then decreases gradually, and it gradually increases
as the progress of the supercycle phase.  This result is in very
good agreement with the PDM analysis (figure \ref{fig:v1504negshvar}).
The same trend is only barely visible in the Fourier analysis.
The strength of the orbital modulation also varies, and can be
seen in the Lasso spectrum as a signal of variable strength,
while the same trend is only barely seen in Fourier spectrum.
Although PDM has an ability to measure the frequency of
a single signal with a high precision, this method cannot
is not designed to deal with a combination of multiple signals.
Although Fourier analysis, on the other hand, can handle
a combination of multiple signals, it has a lower frequency
resolution than in PDM, and its ability to detect the rapid
variation of the frequency is limited.  The Lasso analysis
meets the both requirement (frequency precision and multiple
signals), and our results clearly demonstrate that
the Lasso period analysis is suitable for study of frequency 
variations when multiple signals are present, such as
in positive and negative superhumps in SU UMa stars. 
We also add results for data of Kepler 14th quarter (Q14) recently 
released for public in the last panels of figure
\ref{fig:v1504spec2dlasso} and figure \ref{fig:v344spec2dlasso}.

   There is a free parameter $\lambda$ in giving an $\ell_1$ term
(cf. \cite{kat12perlasso}).  In producing figures 
\ref{fig:v1504spec2dlasso}, \ref{fig:v344spec2dlasso},
we selected $\lambda$ which gives
the best contrast of the signals against the background, i.e.
most physically meaningful parameter.  This $\lambda$ is close
to the most regularized model with a cross-validation error
within one or two standard deviations of the minimum
(cf. R Sct for \cite{kat12perlasso}).
We also applied smearing of the signals between $\pm$3 bins
shifted by 0.5~d considering the width of the window. 

\addtocounter{figure}{-1}
\begin{figure*}
  \begin{center}
    \FigureFile(160mm,120mm){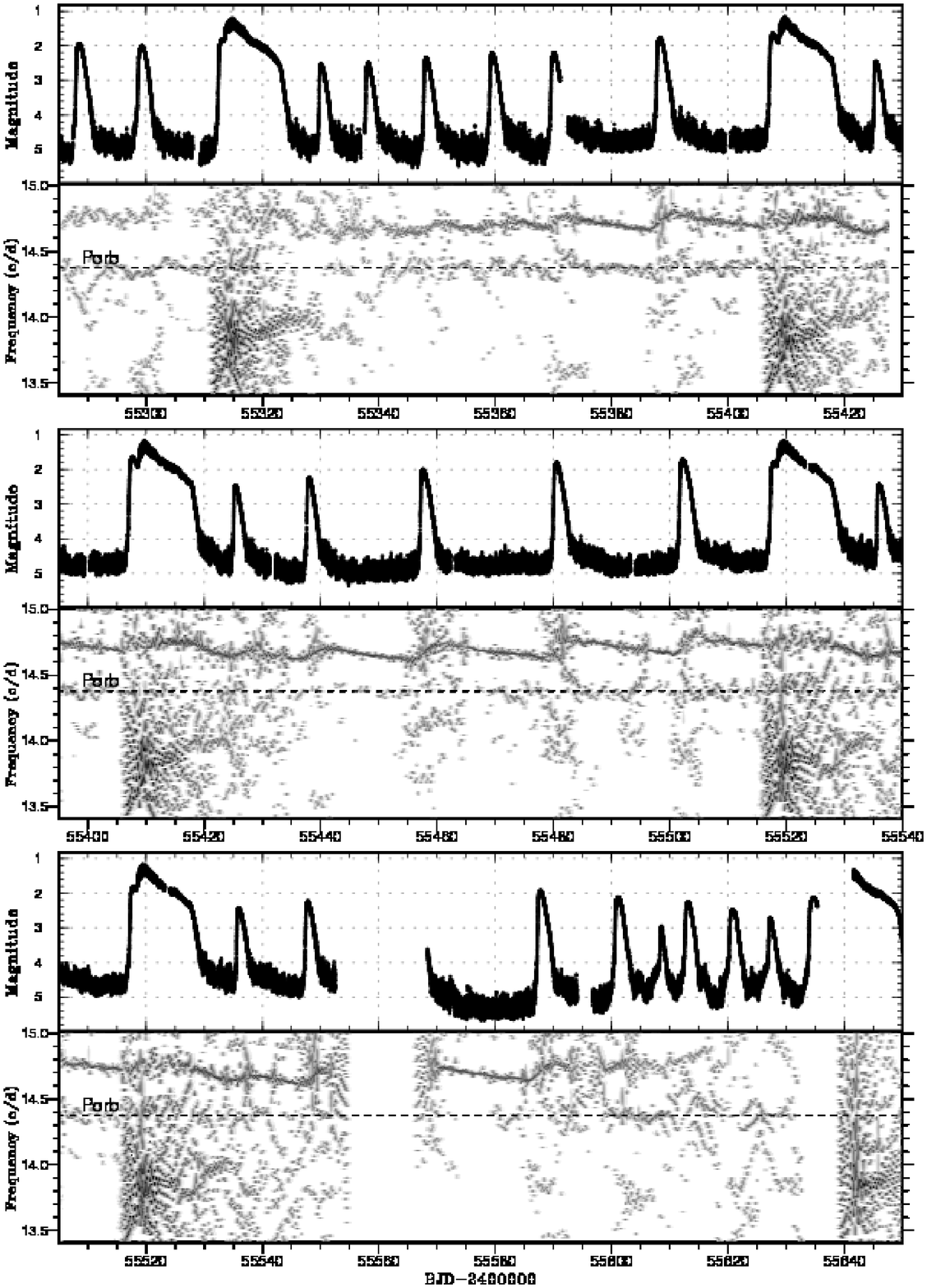}
  \end{center}
  \caption{Two-dimensional lasso power spectrum of the Kepler
  light curve of V1504 Cyg (continued).
  (upper:) light curve; the Kepler data were binned to 0.02~d,
  (lower:) lasso power spectrum ($\log \lambda=-3.9$). The width of 
  the sliding window and the time step used are 5~d and 0.5~d,
  respectively.}
\end{figure*}

\addtocounter{figure}{-1}
\begin{figure*}
  \begin{center}
    \FigureFile(160mm,120mm){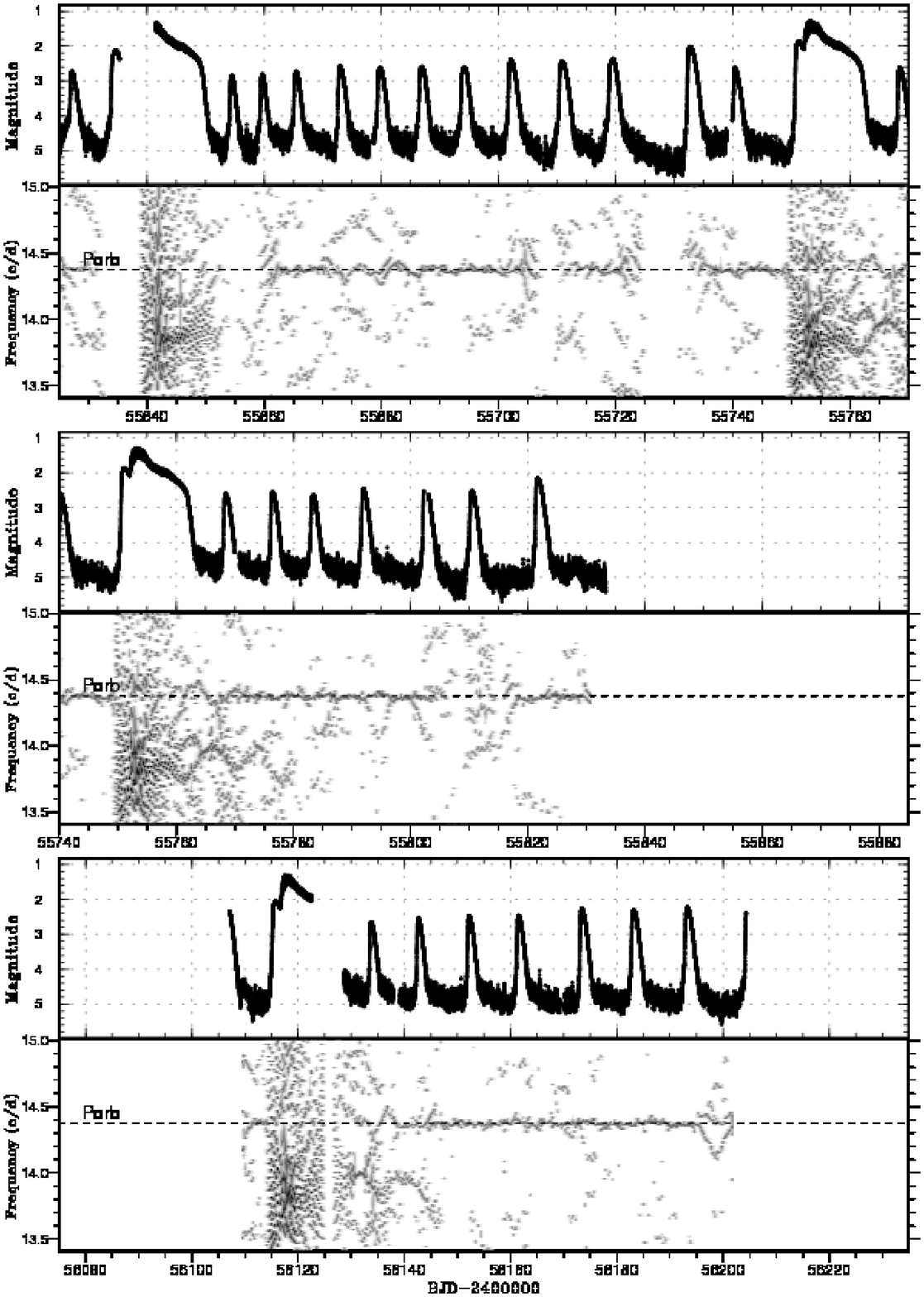}
  \end{center}
  \caption{Two-dimensional lasso power spectrum of the Kepler
  light curve of V1504 Cyg (continued).
  (upper:) light curve; the Kepler data were binned to 0.02~d,
  (lower:) lasso power spectrum ($\log \lambda=-3.9$). The width of 
  the sliding window and the time step used are 5~d and 0.5~d,
  respectively.}
\end{figure*}

\begin{figure*}
  \begin{center}
    \FigureFile(160mm,120mm){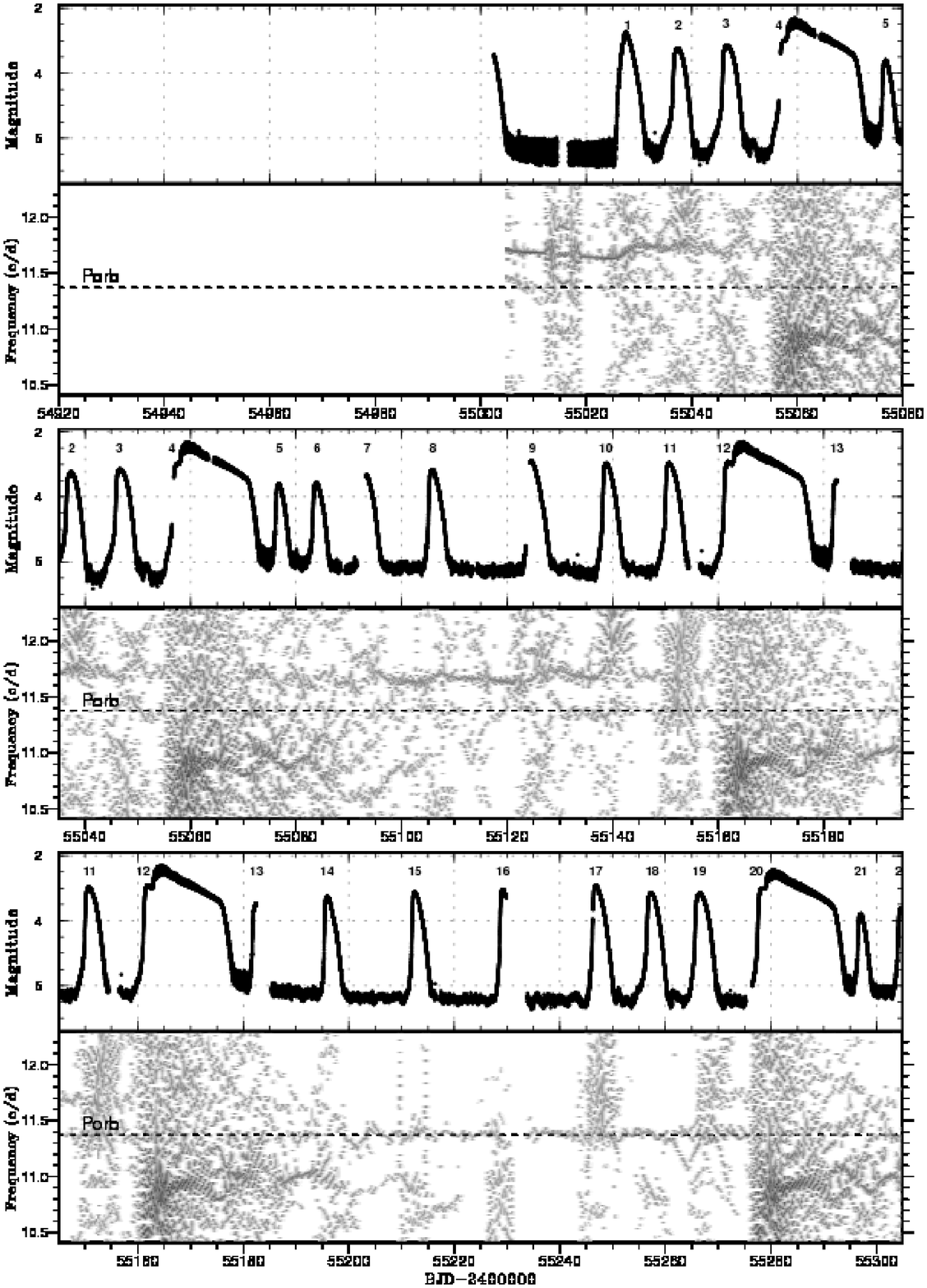}
  \end{center}
  \caption{Two-dimensional lasso power spectrum of the Kepler
  light curve of V344 Lyr.
  (upper:) light curve; the Kepler data were binned to 0.02 d,
  (lower:) lasso power spectrum ($\log \lambda=-4.8$). The width of 
  the sliding window and the time step used are 5 d and 0.5 d,
  respectively.}
  \label{fig:v344spec2dlasso}
\end{figure*}

\addtocounter{figure}{-1}
\begin{figure*}
  \begin{center}
    \FigureFile(160mm,120mm){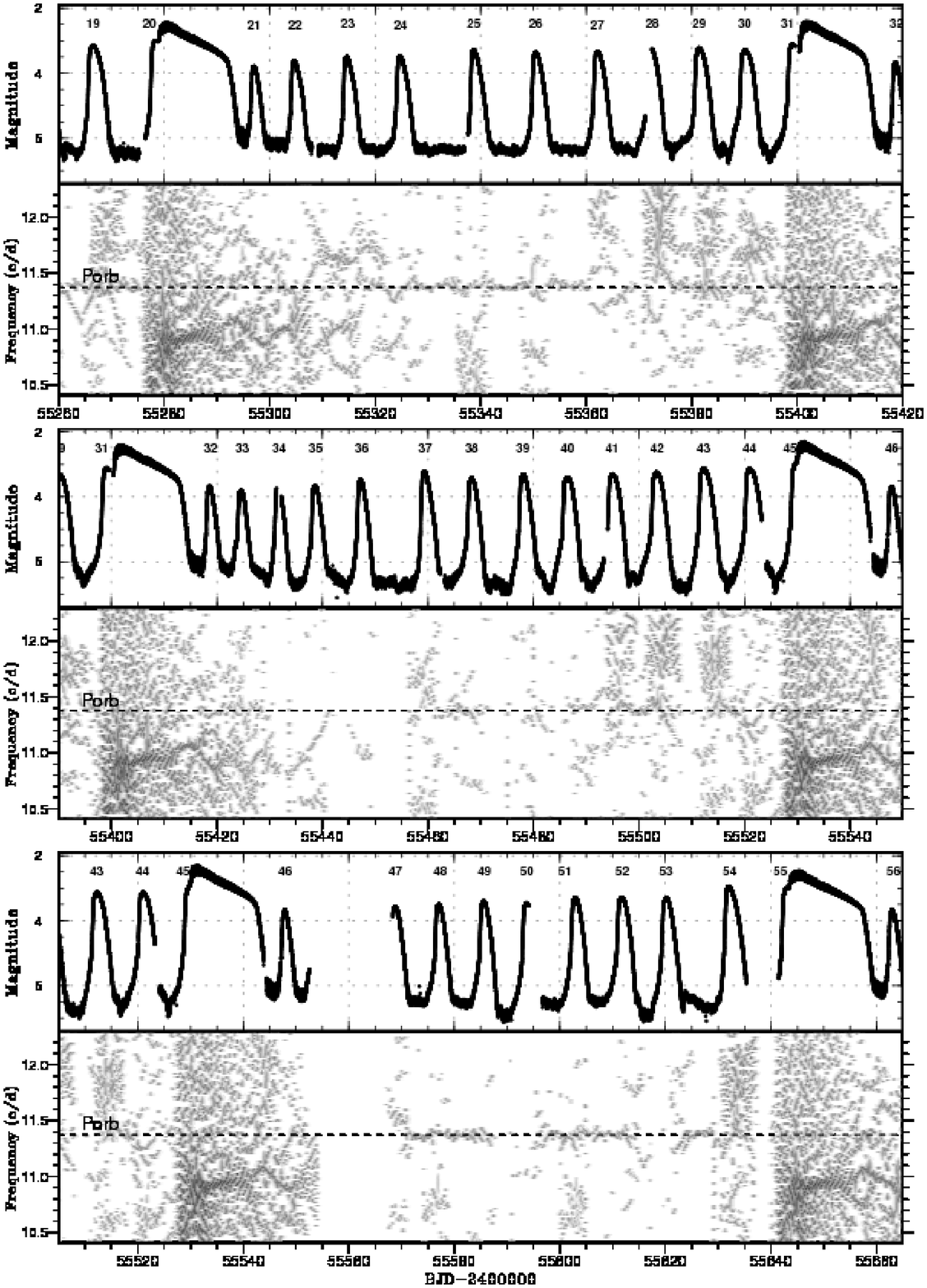}
  \end{center}
  \caption{Two-dimensional lasso power spectrum of the Kepler
  light curve of V344 Lyr (continued).
  (upper:) light curve; the Kepler data were binned to 0.02~d,
  (lower:) lasso power spectrum ($\log \lambda=-4.8$). The width of 
  the sliding window and the time step used are 5~d and 0.5~d,
  respectively.}
\end{figure*}

\addtocounter{figure}{-1}
\begin{figure*}
  \begin{center}
    \FigureFile(160mm,120mm){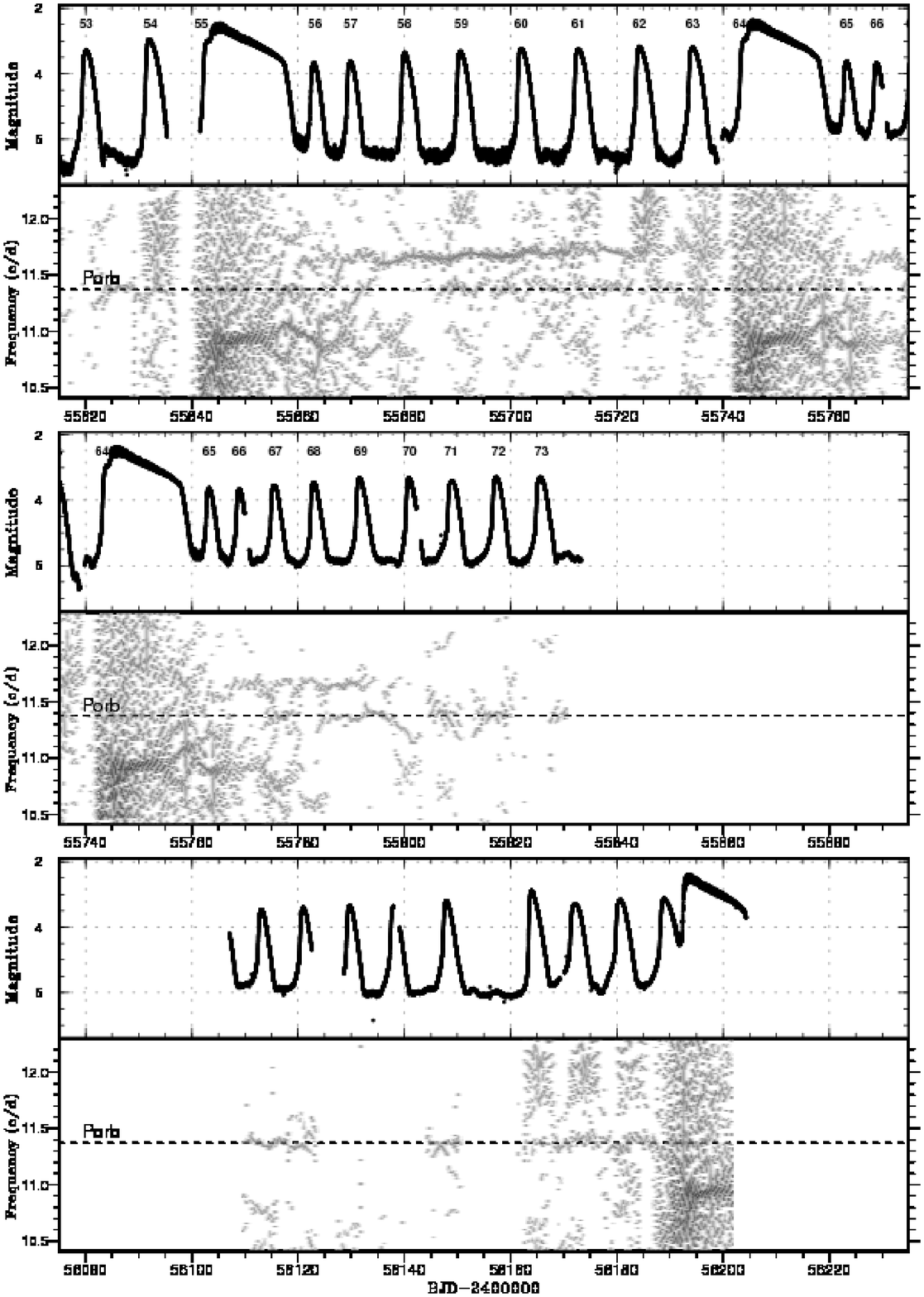}
  \end{center}
  \caption{Two-dimensional lasso power spectrum of the Kepler
  light curve of V344 Lyr (continued).
  (upper:) light curve; the Kepler data were binned to 0.02~d,
  (lower:) lasso power spectrum ($\log \lambda=-4.8$). The width of 
  the sliding window and the time step used are 5~d and 0.5~d,
  respectively.}
\end{figure*}

\subsection{Positive and negative superhumps and their (simultaneous) 
frequency variation}

The positive and negative superhumps of V344 Lyr in Kepler data have already 
been discussed by \citet{sti10v344lyr} and in particular by
\citet{woo11v344lyr} in great details. 

Here we study them from a different 
standpoint, in particular in relation to the supercycle. 
We note here that 
in this section and in the next subsection we deal with the positive superhump 
and the negative superhump as linear eigenmodes of an accretion disk 
(i.e., the eccentric mode and the tilt mode, respectively), and we discuss 
their eigen-frequency variations. We recognize very well that a linear 
approach has some limitation because observed superhumps are quite non-linear. 
In this approach the unperturbed disk, in which these two eigenmodes occur, 
is assumed to be a circular, coplanar Keplerian disk and 
all complications produced by non-linear effects, such as 
density waves, warps, are not taken into account in the unperturbed disk.  
Our approach is to adopt the simplest assumption of linear mode analysis and  
we then examine what this approach tells us. 

As already pointed out by \citet{sti10v344lyr}, the positive superhumps 
which first appeared with the start of a superoutburst continued to be 
seen during the following quiescence and the next normal outburst in V344 Lyr. 
\citet{woo11v344lyr} demonstrated that the positive superhump signal was 
generated by viscous dissipation within the periodically flexing disk 
(i.e., eccentric precessing disk) in the early and main phase of 
superoutburst but in the later phase of superoutburst it was generated 
by impact of gas stream with non-axisymmetric disk rim 
(i.e., that corresponds to so-called late superhump). 

\begin{figure*}
  \begin{center}
    \FigureFile(160mm,100mm){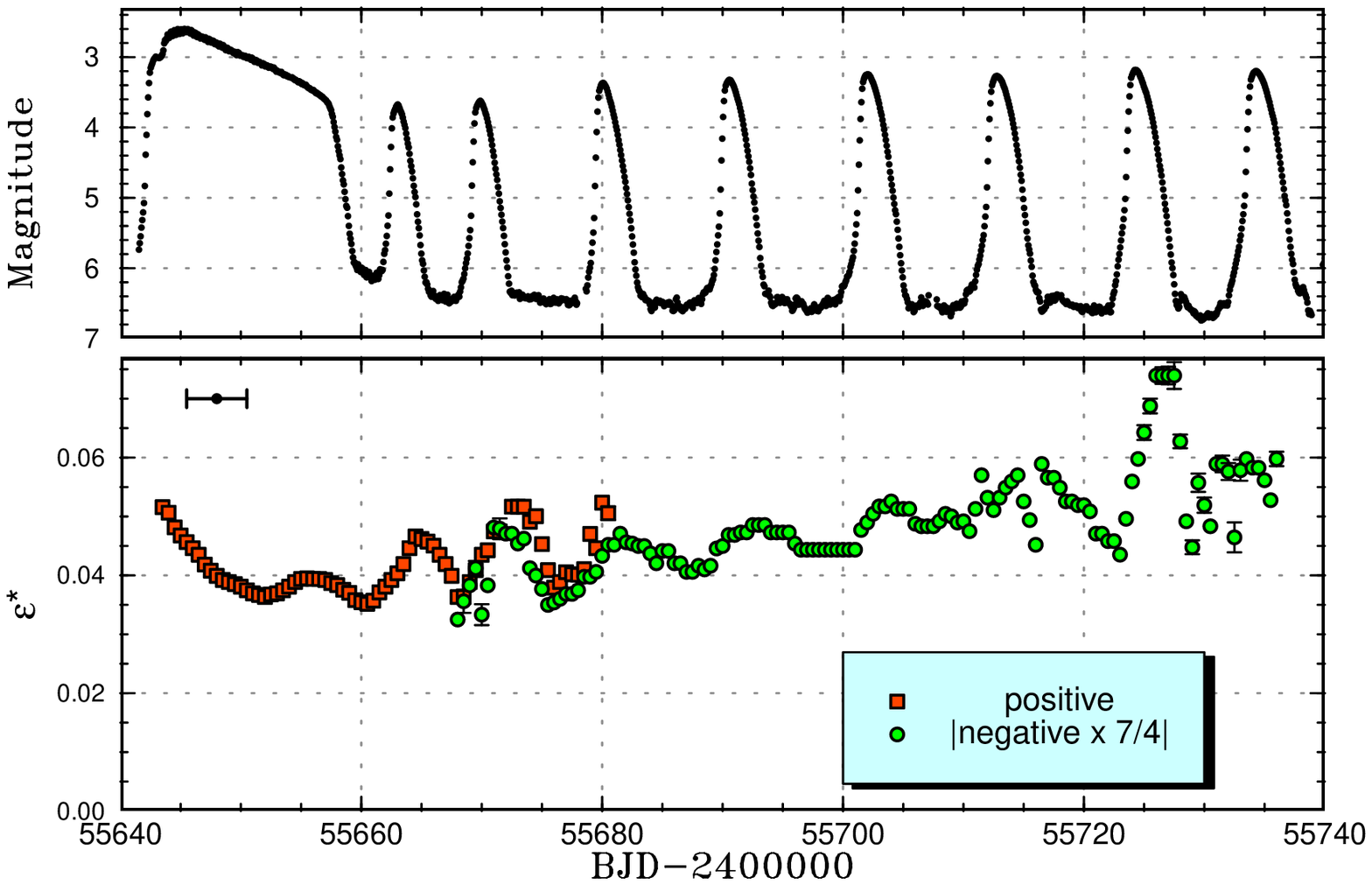}
  \end{center}
  \caption{Variation in precession rates of positive and negative superhumps 
  given by two $\epsilon^*$'s for the supercycle No. 7 of V344 Lyr. 
  (Upper:) Light curve of V344 Lyr for a period of
  BJD 2455640--2455740.  The Kepler data were averaged to 0.07~d bins.
  (Lower:) Absolute values of fractional superhump excesses
  (positive and negative) in frequency scale.
   The window width (5~d) is indicated by a horizontal bar 
   at the upper left corner 
  and the error bars represent 1-$\sigma$ errors in the periods.
  }
  \label{fig:v344posnegper}
\end{figure*}

As seen in figure \ref{fig:v344spec2dlasso} (for superoutbursts No. 1 
and No. 6 of V344 Lyr, see, also figure \ref{fig:v344posnegper}), 
the frequency of 
the positive superhumps exhibits a characteristic variation with progress 
of a superoutburst and the following quiescence and the next normal outburst. 
When the superoutburst is just over and the cooling transition occurs 
at the outer disk-edge and the cooling wave starts to propagate inward, 
the nature of the superhump changes from that of the disk origin to 
that of gas stream as discussed by \citet{woo11v344lyr}.
That is, the late superhump appears and a remnant eccentricity of the disk 
remains for a little while.  In the case of V344 Lyr the remnant eccentricity 
remains during the following quiescence and the next one or 
two normal outbursts.

The positive superhumps are produced by the tidal instability at a strong 
resonance at 3:1 radius when a superoutburst starts (or 
rather the tidal instability and the superhumps initiate a superoutburst)  
and the large amplitude oscillations continue to be seen 
during the superoutburst. In this subsection we concentrate 
our discussions on the late superhump as we discuss the frequency 
variation of the positive superhumps during the superoutburst 
in the next subsection.  

\begin{figure}
  \begin{center}
    \FigureFile(88mm,100mm){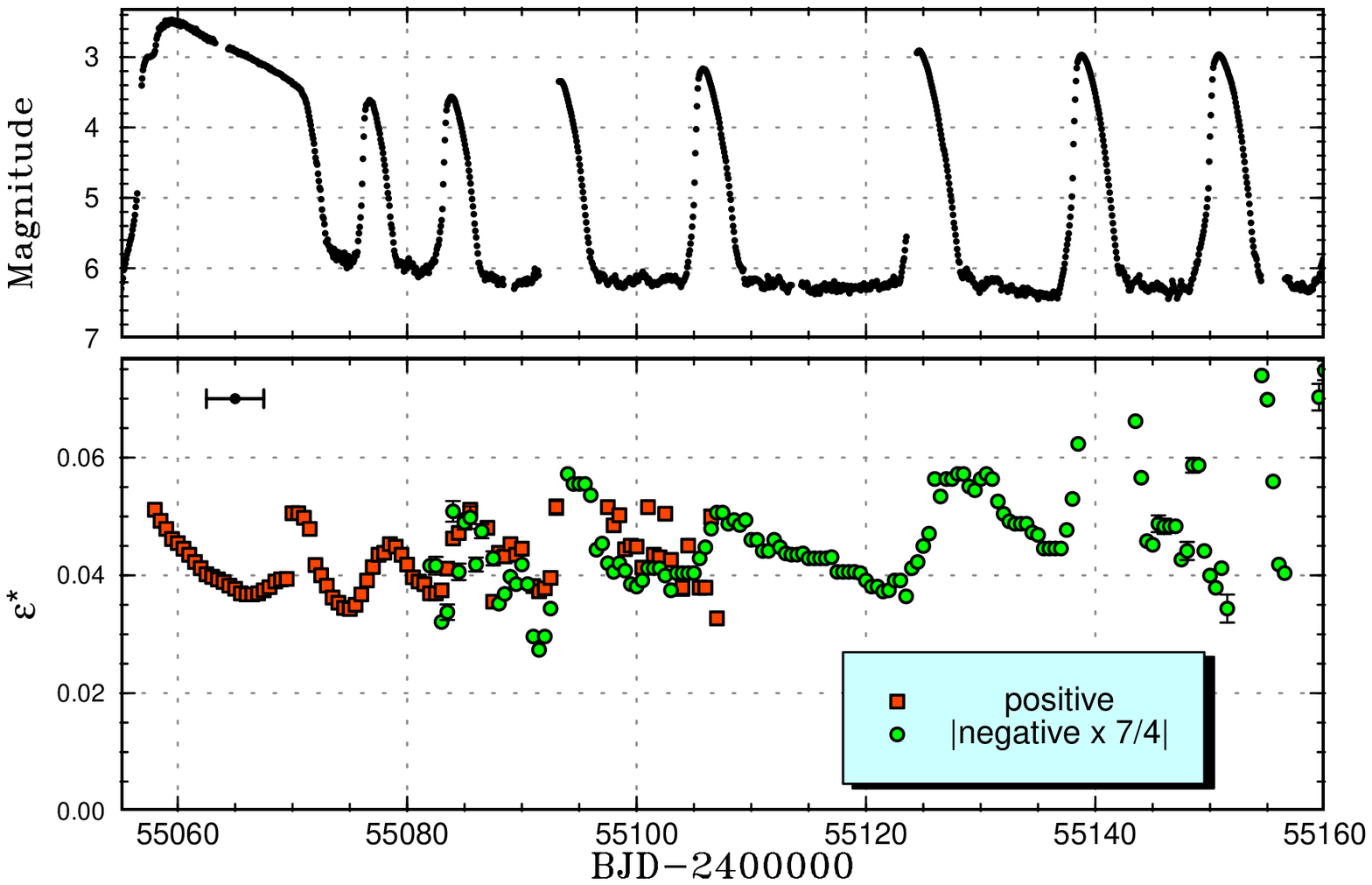}
  \end{center}
  \caption{Variation in precession rates of positive and negative superhumps 
  given by two $\epsilon^*$'s for a supercycle No. 2 of V344 Lyr
  (Upper:) Light curve of V344 Lyr for a period of
  BJD 2455055--2455160.  The Kepler data were averaged to 0.07~d bins.
  (Lower:) Absolute values of fractional superhump excesses
  (positive and negative) in frequency scale.
   The window width (5~d) is indicated by a horizontal bar 
   at the upper left corner 
  and the error bars represent 1-$\sigma$ errors in the periods.
  }
  \label{fig:v344posnegper2}
\end{figure}

\begin{figure}
  \begin{center}
    \FigureFile(88mm,70mm){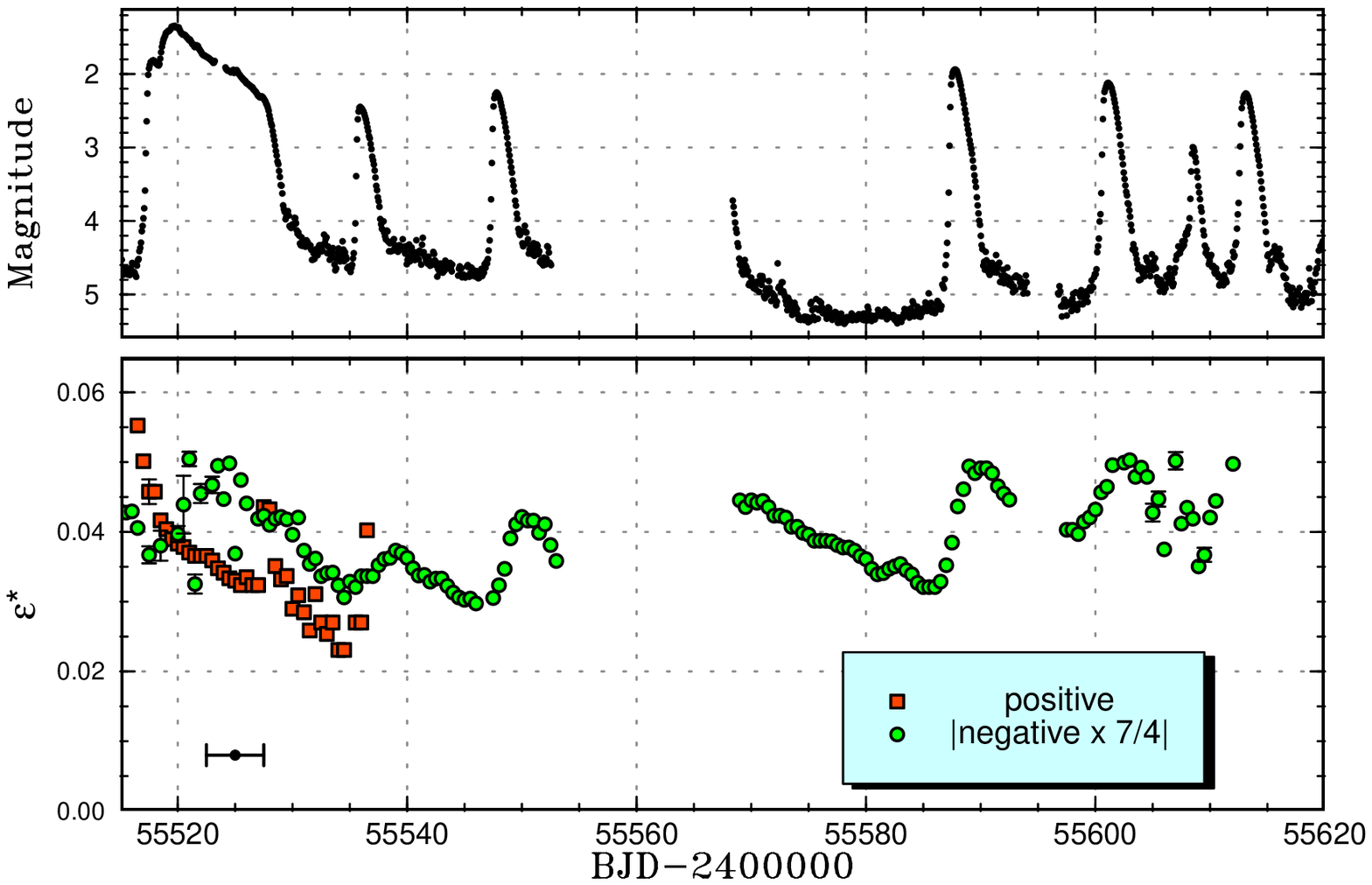}
  \end{center}
  \caption{Variation in precession rates of positive and negative superhumps 
  given by two $\epsilon^*$'s for a supercycle No. 6 of V1504 Cyg.
  (Upper:) Light curve of V1504 Cyg for a period of
  BJD 2455515--2455620.  The Kepler data were averaged to 0.07~d bins.
  (Lower:) Absolute values of fractional superhump excesses
  (positive and negative) in frequency scale.
   The window width (4~d) is indicated by a horizontal bar 
   at the lower left corner 
  and the error bars represent 1-$\sigma$ errors in the periods.
  }
  \label{fig:v1504posnegper3}
\end{figure}

It is usual to define the fractional superhump period excess 
over the orbital period by $\epsilon$, which is written as 
\begin{equation}
\epsilon=(P_{\rm SH}-P_{\rm orb})/P_{\rm orb}.
\label{equ:epsilon}
\end{equation}
The quantity $\epsilon$ is positive for the positive superhump 
(i.e., $\epsilon_+>0$) and it is negative for the negative superhump 
(i.e., $\epsilon_-<0$) by definition.
Since the positive and negative superhump frequency deficiency or excess is 
related to the apsidal precession rate of the eccentric disk 
and the nodal precession rate of the tilted disk, respectively, 
it is convenient to introduce the superhump frequency deficiency 
or excess rate by 
\begin{equation}
\epsilon^*=\frac{\nu_{\rm PR}}{\nu_{\rm orb}}
=\frac{\nu_{\rm orb}-\nu_{\rm SH}}{\nu_{\rm orb}},
\label{equ:epsilonstar}
\end{equation}
where $\nu_{\rm PR}$ is a precession rate of disk, and 
the sign of $\epsilon^*$ is chosen in such a way that $\epsilon^*>0$ 
if the precession is prograde and $\epsilon^*<0$ if it is retrograde. 
These two $\epsilon$'s are related with each other by 

\begin{eqnarray}
\qquad  \epsilon^* =\frac{P_{\rm orb}}{P_{\rm SH}}\epsilon
=\frac{\epsilon}{1+\epsilon} \\ \label{equ:twoepsilon1}
{\rm or} \qquad  \epsilon=\frac{\epsilon^*}{1-\epsilon^*}
\label{equ:twoepsilon2}
\end{eqnarray}

Thus $\epsilon^*$ is slightly smaller than $\epsilon$ in the case of 
the positive superhump while the absolute value of the former 
is slightly larger than that of the latter in the case of the negative 
superhump. 

The precession rate of the eccentric disk, $\nu_{\rm PR}$ in the late 
superhump is basically determined by the dynamical precession of 
the eccentric disk by gravitational field of the secondary star and 
it is given in the lowest order of expansion in the disk radius, $R_d$,  
(see, \cite{osa85SHexcess}) by
\begin{equation}
\epsilon^*_{+}=\frac{\nu_{\rm aPR}}{\nu_{\rm orb}}
=1-\frac{\nu_{\rm pSH}}{\nu_{\rm orb}}
= \frac{3}{4} \frac{q}{\sqrt{1+q}} (\frac{R_d}{A})^{3/2} 
\label{equ:apprecession}
\end{equation}
where the plus sign in the subscript of $\epsilon^*_{+}$ signifies 
the positive superhump and $q=M_2/M_1$ is the mass ratio of the binary, 
$A$ is the binary separation as usual. Here $\nu_{\rm aPR}$ is 
the apsidal precession rate of an eccentric disk. Here we note that 
equation (\ref{equ:apprecession}) is valid for the cold disk 
(i.e., in quiescence) because the pressure effects are unimportant in the 
cold disk and because the eigenfunction of the eccentric mode is confined 
to the outer part of the disk in the case of the cold disk, as will be 
discussed in the next subsection. 

As discussed in Paper I, the precession rate of a tilted disk 
over the orbital frequency is given (see, \cite{lar98XBprecession}) by 
\begin{equation}
\epsilon^*_{-}=\frac{\nu_{\rm nPR}}{\nu_{\rm orb}}
=1-\frac{\nu_{\rm nSH}}{\nu_{\rm orb}}
=-\frac{3}{7} \frac{q}{\sqrt{1+q}} 
(\frac{R_d}{A})^{3/2} \cos \theta, 
\label{equ:noprecession}
\end{equation}
where $\nu_{\rm nPR}$ is the nodal precession rate of a tilted disk, 
and the negative sign signifies its retrograde nature. 
 If we compare two expressions of $\epsilon^*$ for positive and negative 
 superhumps, we find 
\begin{equation}
\frac{\epsilon^*_{+}}{\mid \epsilon^*_{-}\mid }\simeq \frac{7}{4}. 
\label{equ:ratioprecessionrate}
\end{equation}
Here we assumed $\cos \theta \simeq 1$ for a slightly tilted disk.

One of the most interesting aspects in the power spectrum of V344 Lyr is 
that the positive superhump signal (the late superhump) is seen together 
with the negative superhump signal during quiescence just after a  
superoutburst and the next two normal outbursts in supercycle No. 7, 
i.e., in a period between days BJD 2455670 to 2455680. 
This means that the disk was eccentric and tilted simultaneously. A question 
then naturally arises how these two different superhumps are generated 
simultaneously because 
both light variations of the late superhump and of the negative superhump 
in quiescence are thought to be produced by the impact of gas stream. 
Although the exact mechanism is not known yet, it is most likely 
that a part of gas stream left from the Lagrangian point collides 
with the rim of an eccentric disk, producing the positive superhump signal 
while other part of gas stream spills over the rim and arrives in the inner 
part of a tilted disk, producing the negative superhump signal. In fact, 
\citet{mon12negposSH} has found that the positive and negative superhumps 
appear simultaneously in her numerical simulations. 

In paper I we have demonstrated that the frequency 
variation of the negative superhump is a good indicator of the disk-radius 
variation and its variation thus obtained fits very well with a prediction of 
the thermal-tidal instability model. It will be very interesting if 
we analyze frequency variations of the two different superhump signals. 
We thus study the frequency variations of positive superhumps 
and negative superhumps during supercycle No. 7 in V344 Lyr 
for a period from day BJD 2455640 to 2455740. 

To do so, we have used a sliding window with a width of 2~d and time step 
of 0.5~d and we have calculated frequencies 
by using PDM in obtaining the periods. 
We illustrate the frequency deficiency (or excess) over the orbital 
frequency for the negative and positive superhumps thus obtained in 
figure \ref{fig:v344posnegper}
simultaneously where the results, $\epsilon^*_+$, 
for the positive superhumps are shown in filled squares and 
those for the negative superhumps are shown in filled circles
by converting them to those of $(7/4)\times \mid \epsilon^*_- \mid$ 
in order to show within one figure. 

Let us first look at the variation in frequency excess of 
the negative superhumps $\mid \epsilon^*_- \mid$ 
in figure \ref{fig:v344posnegper}. 
The absolute value of $\epsilon^*_-$ increases 
whenever a normal outburst occurs and it decreases during quiescence. 
Its mean level in one normal outburst cycle increases 
with advance of supercycle phase which is in a good agreement with that of 
V1504 Cyg. One exception is that occurring in the second normal 
outburst from the last. This is due to the impulsive negative humps 
which will be discussed separately in subsection \ref{sec:impulsivensh}
below.

The variation in $\epsilon^*_+$ for the late superhump in quiescence 
and the first and the second next normal outbursts just 
after a superoutburst No. 6 exhibits a same pattern with that of 
negative superhump, that is, an increase when a normal outburst occurs 
and it decreases during quiescence. When these two 
oscillations overlap, two $\epsilon^*$ values agree fairly well 
with each other if converted by equation (\ref{equ:ratioprecessionrate}) 
as seen in figure \ref{fig:v344posnegper}. 

We think that an agreement of these two variations 
is not just chance coincidence but it demonstrates that  both the frequency 
variations of the late superhump and of the negative superhumps represent 
the disk-radius variation and it strengthens our conclusion of Paper I, 
supporting the thermal-tidal instability model.  

Figure \ref{fig:v344posnegper2} illustrates the same type of figure but for 
the supercycle No. 2 of V344 Lyr, basically confirming the above mentioned 
results. Figure \ref{fig:v1504posnegper3} is another example for 
the supercycle No. 6 of V1504 Cyg, showing a similar pattern. 
However, points based on the nSH are systematically higher 
than those based on the pSH in an interval BJD 2455530--2455535 
and this point will be discussed in the Appendix.

 So far we have used equation (\ref{equ:noprecession}) for the nodal 
 precession rate of a tilted accretion disk which was obtained 
 by \citet{lar98XBprecession} and where a proportionality constant appearing 
 in equation (\ref{equ:noprecession}) is 3/7. 
However different authors derived slightly different expressions for this 
coefficient because of slightly different assumptions. For instance, 
\citet{mon09diskprecession} obtained 15/32 instead of 3/7 for this coefficient 
[see, equation (38) of \citet{mon09diskprecession}]. Montgomery's 
expression is larger by a factor 1.09 than 3/7. 

In order to take this effect into 
account, here we introduce a correction factor $\eta$ to allow 
for a different expression of the nodal precession rate in such a way 
that the numerical coefficient in equation (\ref{equ:noprecession}) is 
modified as  
\begin{equation}
\frac{\nu_{\rm nPR}}{\nu_{\rm orb}}
=-\eta\frac{3}{7} \frac{q}{\sqrt{1+q}} 
(\frac{R_d}{A})^{3/2} \cos \theta, 
\label{equ:noprecession_mod}
\end{equation}

Then we have $\epsilon^*_+/\mid \epsilon^* _{-}\mid \simeq \eta^{-1}7/4$
for equation (\ref{equ:ratioprecessionrate}). 
In the Appendix  we will discuss a possible value for $\eta$ by 
taking into account for a different mass distribution in the disk. As shown 
in the appendix, difference of the correction factor $\eta$ from 1 is rather 
small, typically less than 10\% but it could be as large as 20\% 
in some cases. 

\subsection{Frequency Variations of the Positive Superhumps 
during Superoutbursts}

The interpretation of the frequency (or period) variations of the positive 
superhumps during superoutbursts is much difficult and so we discuss them 
separately from those in quiescence and in normal outbursts. Extensive survey 
of period variations of superhumps in SU UMa stars in a form of the 
$O-C$ diagram have been accumulated in a series of papers (I--IV) 
by \citet{Pdot}, \citet{Pdot2}, \citet{Pdot3}, \citet{Pdot4}.
Data based on the Kepler light curves of V1504 Cyg and V344 Lyr supplement 
these observations in particular for SU UMa stars 
of longer orbital periods and of high mass transfer systems. 

\subsubsection{Factors Contributing to the Frequency Variations 
of the Positive Superhumps}

The superhump periodicity (and hence frequency) of the positive superhump 
is produced by a synodic period between a precessing eccentric disk and 
the orbiting secondary star, which is given by  
$\nu_{\rm pSH}=\nu_{orb}-\nu_{\rm aPR}$. Theoretical precession rate 
of an eccentric disk in relations with superhumps of SU UMa stars has been 
discussed by \citet{osa85SHexcess}, \citet{hir90SHexcess}, \citet{lub91SHa},
\citet{hir93SHperiod}, and 
\citet{goo06SH}  among others.  
\citet{lub92SH} showed that 
the precession rate of eccentric disk consisted of three different terms,  
which is written by 
\begin{equation}
\nu_{\rm aPR}=\nu_{\rm dyn}+\nu_{\rm pressure}+\nu_{\rm stress},
\label{equ:Lubows}
\end{equation}
where the first term,$\nu_{\rm dyn}$, represents a contribution 
to disk precession due to tidal perturbing force of the secondary, 
giving rise to prograde precession, the second term, $\nu_{\rm pressure}$, 
does that due to pressure effect (or an effect of finite thickness of 
the disk), giving rise to retrograde precession, and the last term, 
$\nu_{\rm stress}$, does that due to wave-wave interaction, 
which arises when the eccentric mode grows or decays 
in time, giving rise to either retrograde or prograde precession. 
The third term may become important when eccentricity (and thus superhump) 
grows rapidly in the initial stage of superoutburst or 
when it decays in the final stage of a superoutburst, but \citet{lub92SH} 
argued that it is much smaller than the first two terms. In what follows, 
we concentrate our discussion to the first two terms.  

\subsubsection{Eigenfrequency Expression of the Frequencies of
the Positive Superhumps}

\begin{figure*}
  \begin{center}
    \FigureFile(140mm,100mm){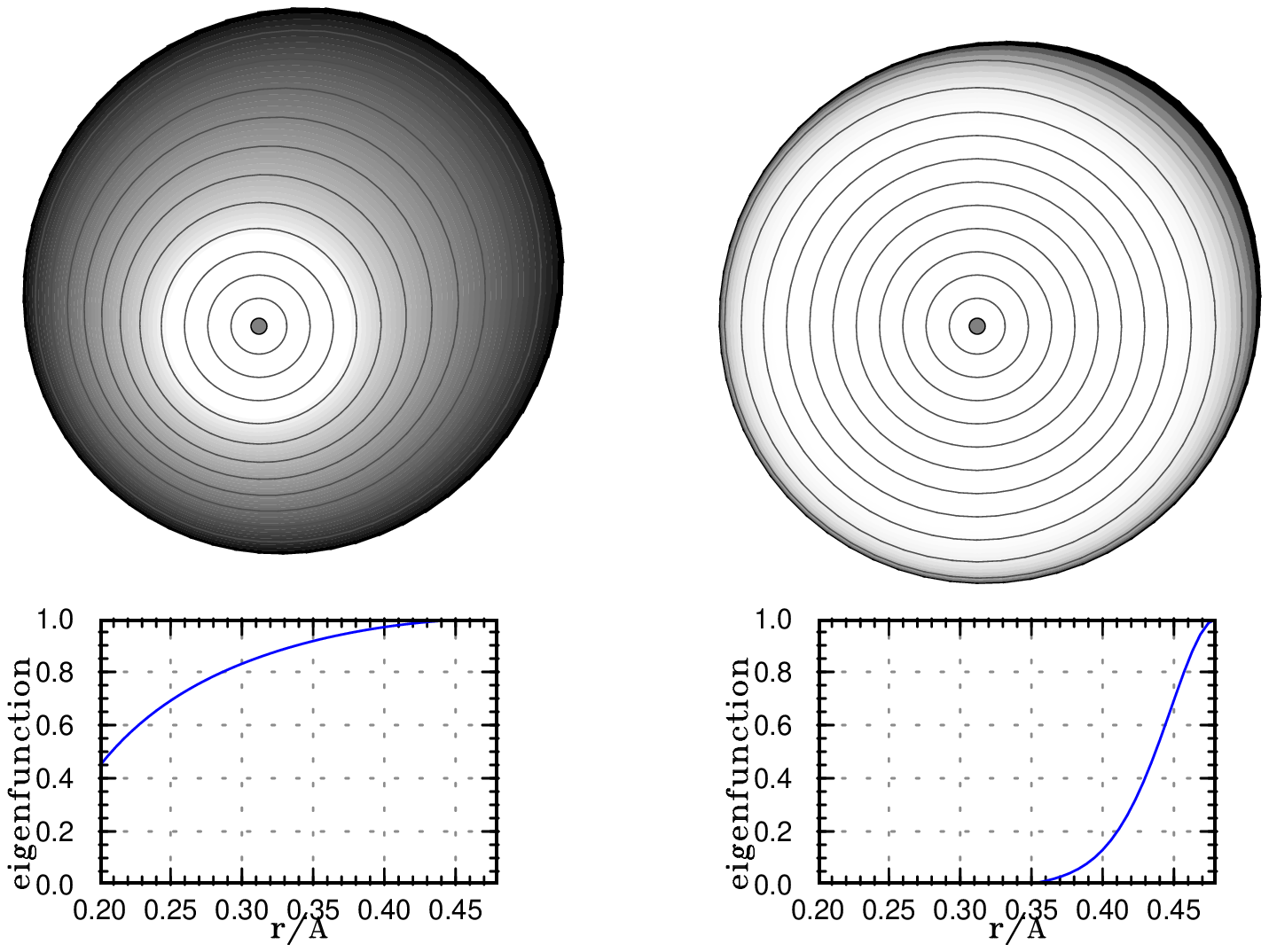}
  \end{center}
  \caption{Schematic illustration of eigenfunctions of the superhump mode 
  $m=1$.
  Darker colors represent larger values in eigenfunction and larger
  eccentricities (displacement).  
  The eccentricity shown here is important only in relative sense with 
  different radii and its absolute value has no meaning because of linear 
  mode analysis.
  The eigenfunctions of these figures are for the case of $q=0.15$.
  (Left:) hot disk ($c_0 = 0.1$) where $c_0$ is the dimensionless 
  sound velocity.  The eigenfunction takes large values
  for a large range of the disk radius, and the eccentricity wave
  is present in a wider region of the disk.
  ($\omega=0.0289$ where $\omega$ is the dimensionless eigenfrequency 
  which is equivalent to $\epsilon^*_{+}$) .
  (Right:) cold disk ($c_0 = 0.005$).  The eigenfunction takes large values
  only near the outer edge of the disk, and the eccentricity wave
  is trapped in the outermost region of the disk. ($\omega=0.0492$).
  In the left (hot) case, a larger portion of the disk becomes
  elliptic and the precession rate of the entire disk is smaller
  due to the contribution of slower local precession rates
  in the inner radii.
  }
  \label{fig:eigenpict}
\end{figure*}

The precession rate of eccentric disk was discussed in terms of an 
eigenfrequency of an eccentric mode in an accretion disk 
(the fundamental mode of $m=1$ in a thin disk, which 
hereafter we call the superhump mode)  
by \citet{hir93SHperiod} and \citet{goo06SH}. 
\citet{hir93SHperiod} treated this as a free disk mode 
while \citet{goo06SH} did it as a complex eigenmode so that growth 
and decay of the mode could be discussed together with frequency. 
Since we are here interested only in frequency of eccentric mode 
and since we do not go into the problem of its excitation, 
we adopt discussion by \citet{hir93SHperiod} in what follows. 

As shown by equation (32) in \citet{hir93SHperiod}, the eigenfrequency 
of $m=1$ mode is written in a variational form as 
\begin{equation}
\nu_{\rm aPR}=\frac{\int^{R_d} _{r_1} \nu_{\rm pr} (r) g(r) 
\mid X(r) \mid ^2 dr
-\int^{R_d} _{r_1} f(r) \mid dX(r)/dr \mid ^2 dr} 
{\int^{R_d} _{r_1} g(r) \mid X(r) \mid ^2 dr}
\label{equ:variationalform}
\end{equation} 
where $r$, $r_i$, and $R_d$ are radial distance from the central star in the 
disk, that of the inner edge, and of the outer edge of the disk, 
respectively,  
$X(r)$ is an eigenfunction of the superhump mode
(see figure \ref{fig:eigenpict}), 
and $g(r)$ is a weighting function which increases with radius $r$, 
while $\nu_{\rm pr} (r)$ is the local precession rate of a ring with radius 
$r$ which is given by 
\begin{equation}
\nu_{\rm pr} (r)/\nu_{\rm orb}=\frac{3}{4} \frac{q}{\sqrt{1+q}}(r/A)^{3/2}
\end{equation}
and $f(r)$ is a function proportional to square of local sound speed, 
(i.e., local temperature). 

It is obvious that the first and the second terms in the numerator 
of equation (\ref{equ:variationalform}) 
represent the dynamical effect and the pressure effect in 
equation (\ref{equ:Lubows}), respectively. As for the dynamical effect, 
if we neglect for a moment the second integral 
in the numerator of equation (\ref{equ:variationalform}), we find that 
the eigenfrequency of the superhump mode is given by a mean of the local 
precession rate of rings in the disk with an appropriate weight determined 
by the eigenfunction. The pressure effect (or an effect 
due to the finite thickness of the disk) affects on the eigenfrequency 
of the superhump mode in two ways. It firstly gives rise to retrograde 
precession through the second term in the numerator of equation 
(\ref{equ:variationalform}) and there is another indirect effect 
which comes from the first term in the same equation 
through the weighting function. The weighting function is sensitive 
to the eigenfunction which is in turn very sensitive to the disk temperature. 
These two effects work for reduction of the precession rate in a sense 
that higher the disk temperature (figure \ref{fig:eigenpict} left),
the lower the eigenfrequency (i.e., the precession rate).
As a matter of fact, the second effect dominates over the first effect 
as demonstrated by \citet{hir93SHperiod}.

If the disk is cold, the eigenfunction is 
well confined in the outermost region of the disk
(figure \ref{fig:eigenpict} right) and thus the frequency is 
given by the local precession rate of the outermost part of the disk, 
which justifies equation (\ref{equ:apprecession}) for pSH in quiescence. 
(Here we note once more that our treatment is of linear mode analysis in which 
the unperturbed disk is assumed to be a circular, coplanar, Keplerian disk 
and any complicated phenomena which may arise by non-linear effects such as 
spiral density waves are not taken into account.)   
In the previous subsection 
we have demonstrated that frequencies of the negative superhump 
and the positive superhump vary in unison in quiescence and in normal 
outbursts when they are represented in terms of $\epsilon^*$.

To summarize briefly, the frequency of positive superhumps is reduced 
when the disk is hot and the superhump oscillation spreads to a large
range of the radius (large eigenfunction even in small disk radii).
To the contrary, the frequency of positive superhumps can take a 
large value when the superhump oscillation is confined
to the outer most region, either when the disk is cold or when the superhump
oscillation has not enough time to spread immediately after its
excitation.

\subsubsection{Frequency Variation of Positive Superhumps during
Superoutburst: Observations}

Let us now examine observed variations in frequency (or period) of pSH 
during the superoutburst. As seen in figure \ref{fig:v344posnegper}, 
the quantity $\epsilon^*$ 
representing the disk precession rate (or fractional excess of
the positive superhump period) 
starts from its highest value at the start of the superoutburst 
but it decreases very rapidly in initial phase but later more slowly, 
approaching to a constant small value during the main part of 
the superoutburst. As the end of the superoutburst 
draws near, the nature of the superhump changes from the ordinary superhump 
to the late superhump, $\epsilon^*$ increases to a new local maximum. 
When the plateau stage of the superoutburst ends 
and the rapid decay of light curve starts, $\epsilon^*$ 
decreases continuously 
to quiescence. If the remnant eccentricity survives further, $\epsilon^*$ 
decreases in quiescence and it increases when the next outburst occurs.
Our results basically confirm the period variation of the positive superhump 
found in V344 Lyr by \citet{woo11v344lyr}. 

\subsubsection{Frequency Variation of Positive Superhumps during
Superoutburst: Interpretation}

We now present our interpretation of observed variation in the precession rate,
$\epsilon^*$. Our interpretation is still in a speculative state 
and more observations and theoretical investigation are needed to clarify 
the variation of superhump periodicity. 
 Since our discussion is confined only to qualitative one, 
we discuss it sometime in terms of the positive superhump period 
and sometime in terms of the disk precession rate, $\epsilon^*$. 
We first note that the disk radius variation and the pressure effect 
discussed above are two major causes responsible for the period variation of 
the positive superhump. The main objective of this subsection is to clarify 
how these two effects make an interplay.

We interpret that the highest value of $\epsilon^*$ at the start 
of the superoutburst (i.e., the stage A of superhump in Kato et al.'s 
classification) is most likely given by that corresponding 
to the precession rate at the 3:1 resonance radius in the outer edge. 
Its following rapid decrease in the SH period is then understood as due 
to propagation of the eccentricity wave to the inner 
part of the disk by which a larger portion of the disk is involved to 
determine the over-all precession rate of the disk. That is, the pressure 
effects have decreased the SH period appreciably. After the superoutburst 
reaches its maximum, the superhump period (and so the precession rate 
$\epsilon^*$) decreases with much slower pace toward the end of 
the superoutburst. This slow decrease in the pSH period is understood as due 
to a slow decrease in the disk radius as the mass and angular momentum of 
the disk are depleted during the plateau stage of the superoutburst by 
mass accretion and strong tidal removal of angular momentum of the disk 
as the thermal-tidal model indicates. 
As the superoutburst approaches to its end, the disk 
begins to cool and it brings about an ease of the pressure effects 
which makes  
the eigenfunction of the superhump mode to be more concentrated in the outer 
part.  This in turn brings about a rapid increase in the superhump 
period when the late superhump starts. Once the cooling transition occurs 
at the outer edge of the disk, the cooling front propagates inward, 
extinguishing the superoutburst. The period variation now becomes simple, 
basically due to the disk radius variation in the cold state 
as discussed in the preceding subsection.
The variation of pSH period during the superoutburst can be understood 
qualitatively in the framework of the thermal-tidal instability model.

\subsubsection{Change of the Disk Radius during a Supercycle as Seen
from the Variations of the Negative Superhump Periods}

\begin{figure*}
  \begin{center}
    \FigureFile(160mm,100mm){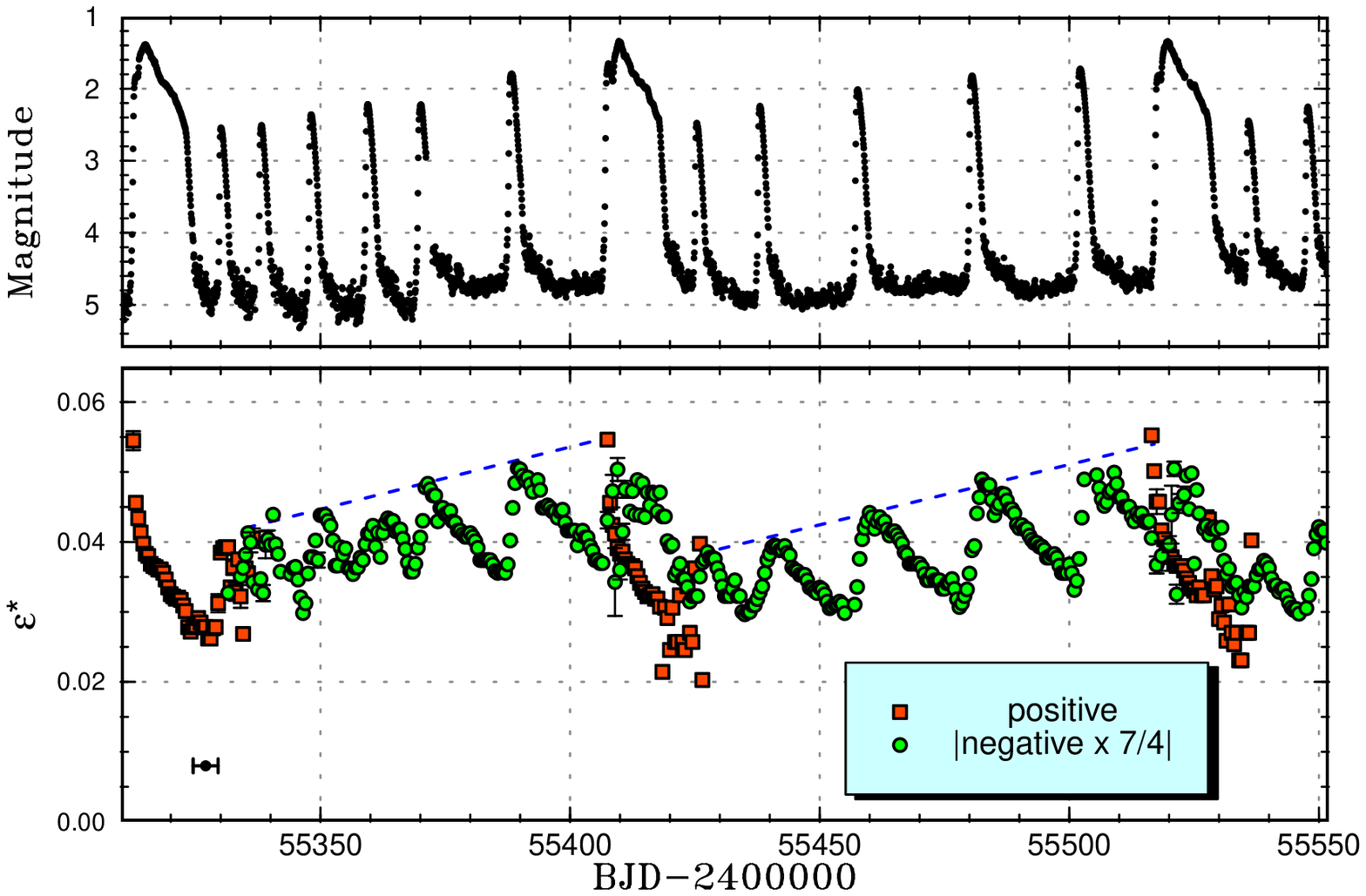}
  \end{center}
  \caption{Variation in precession rates of positive and negative superhumps 
 given by two $\epsilon^*$'s for complete supercycles No. 4 and 5 
 of V1504 Cyg.
  (Upper:) Light curve of V1504 Cyg for a period of
  BJD 2455310--2455520.  The Kepler data were averaged to 0.07~d bins.
  (Lower:) Absolute values of fractional superhump excesses
  (positive and negative) in frequency scale.
  The broken lines represent the trend of the value of
  $\mid \epsilon^*_{-}\mid \times 7/4$ at the peak of 
  every normal outburst and its extrapolation to
  the time of the precursor peak of the next superoutburst.
   The window width (4~d) is indicated by a horizontal bar 
   at the lower left corner 
  and the error bars represent 1-$\sigma$ errors in the periods.
  }
  \label{fig:v1504posnegper12}
\end{figure*}

This general picture can be confirmed from figure \ref{fig:v1504posnegper12} 
which exhibits variations in the two $\epsilon^*$ and which 
covers two supercycles No. 4 and 5 of V1504 Cyg. We first note that  
the variations in the nSH frequency in the supercycle No. 4 
(from BJD 2455325 to BJD 2455406) 
represented by $\mid \epsilon^*_{-}\mid \times 7/4$ 
show a secular increase in disk radius with the advance of supercycle phase 
upon which superimposed are a sudden increase in the disk radius 
when a normal outburst occurs and its gradual decrease during quiescence. 
If we connect the value of $\mid \epsilon^*_{-}\mid \times 7/4$ at 
the peak of every normal outburst by a spline curve 
and extrapolate it to the time of the precursor peak, 
we find this value is about 0.055 (broken lines in
figure \ref{fig:v1504posnegper12}). If we take into account 
the correction factor $\eta\simeq 0.94$ for the hot disk, 
this value becomes about 0.058. 
In either case, its estimated 
value agrees fairly well with the value of $\epsilon^*_{+}$ for the positive 
SH at the start of the superoutburst No. 4. We suggest that these two values 
represent two precession rates corresponding to the 3:1 resonance disk radius.

\subsubsection{Separation of the Variable Radius and Pressure Effect}

The frequency of the negative superhump decreases monotonically during 
the superoutburst No. 4 (BJD 2455406--2455419) as seen both in 
figure \ref{fig:v1504negshvar}
and figure \ref{fig:v1504posnegper12}. This indicates a monotonic decrease 
in the disk radius because the nodal precession rate of a tilted disk depends 
only on the disk radius during the hot viscous plateau stage 
as the density distribution is self-similar as discussed 
in the Appendix and the correction factor discussed 
in equation (\ref{equ:noprecession_mod}) is $\eta\simeq 0.94$  

We now compare these two $\epsilon^*$ during the superoutburst No. 4 in 
figure \ref{fig:v1504posnegper12}. The values shown in red square 
corresponding to the pSH are systematically smaller than those shown 
in green filled-circle corresponding to nSH, roughly speaking about 30\%. 
This difference is due to the pressure effects discussed above. 
By this way we can now separate the pressure effects on 
the positive superhump periodicity from that of the disk radius variation 
during a superoutburst if the positive SH and negative SH signals coexist. 
This basically confirms our interpretion presented above 
on the frequency variation of the positive superhump during a superoutburst. 

\subsection{Impulsive (failed) negative superhumps}\label{sec:impulsivensh}

\begin{figure*}
  \begin{center}
    \FigureFile(88mm,110mm){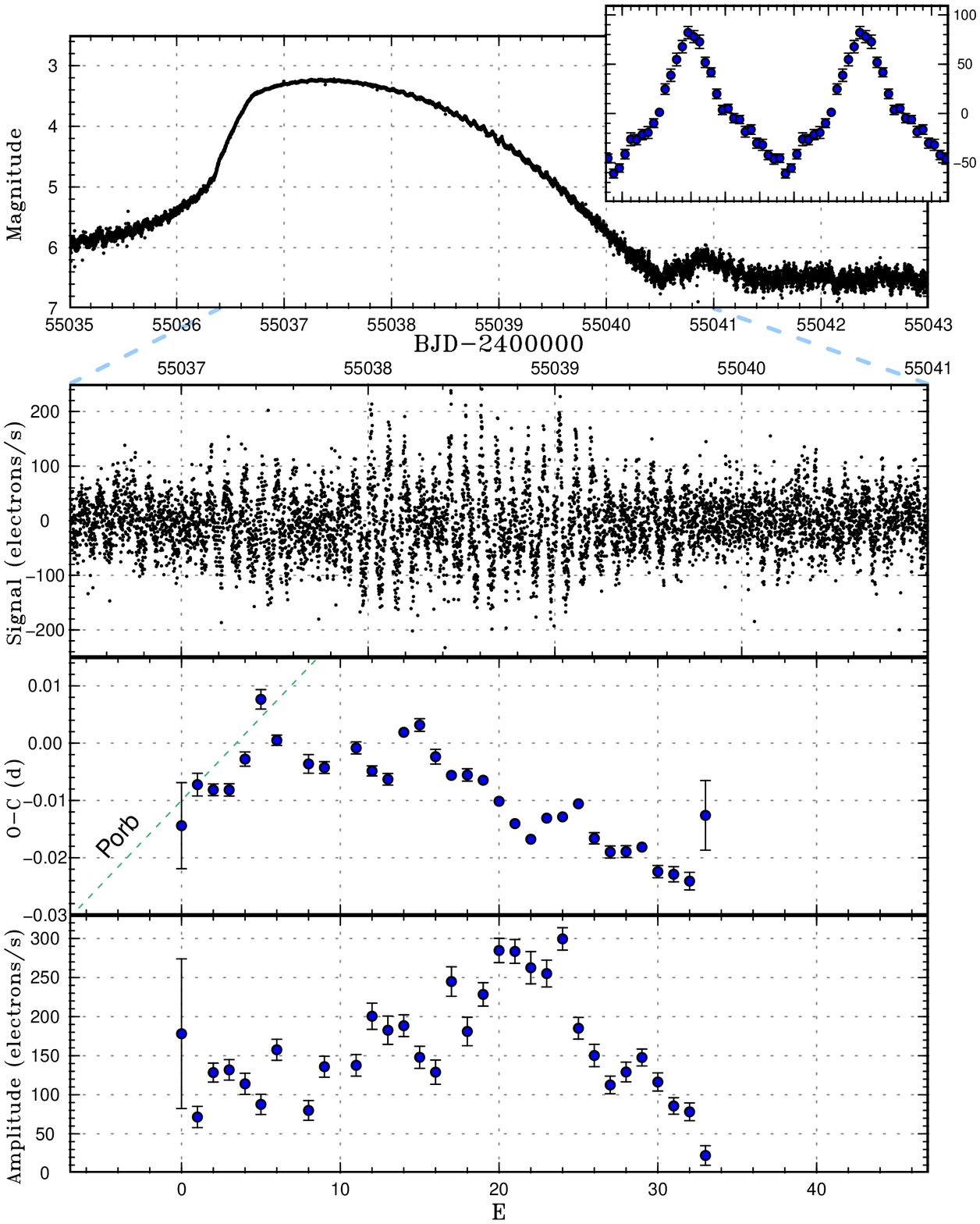}
    \FigureFile(88mm,110mm){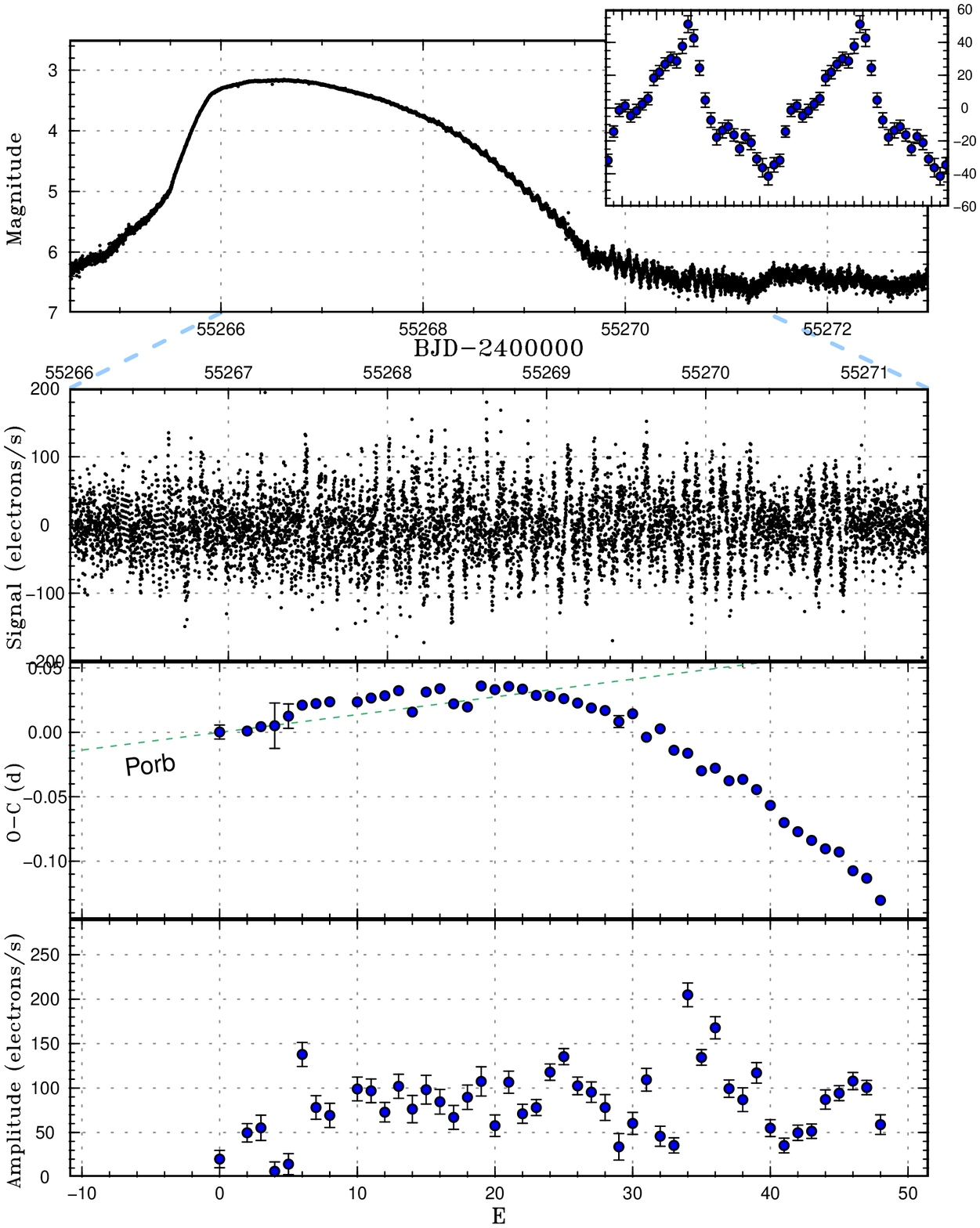}
  \end{center}
  \caption{Two examples of failed negative superhumps in 
  normal outburst (outbursts No. 2 and 19).
  The panels from top to bottom (each outburst) are global light curve,
  pulsed flux, $O-C$ diagram and amplitudes of failed negative
  superhumps.  At the upper right corner, an inset is placed to show
  the mean profile of the failed negative superhumps
  [averaged profile for BJD 2455038--2455040 (No. 2) and
  BJD 2455267.6--2455269.6 (No. 19) when negative superhumps
  were most stably seen.]
  The periods used for drawing the $O-C$ diagrams were 0.085~d for
  outburst 2 and 0.086527~d for outburst 19 (cf. the orbital
  period is 0.087903~d, drawn with dashes in the $O-C$ diagrams.).
  }
  \label{fig:v344failnsh}
\end{figure*}

One of the most important outstanding problems in V344 Lyr and V1504 Cyg 
is the problem of origin of the negative superhump, in other words, 
a problem of the origin of the disk tilt, how the disk tilt 
is produced and how it is maintained. 

Observationally the negative superhump in V1504 Cyg first appeared around 
BJD 2455280 and its amplitude grew until the middle of SC No. 4 
(around BJD 2455350) and then it kept its large amplitude over 200~d 
until the late stage of SC No. 6 and it finally disappeared 
around BJD 2455610 before the next superoutburst. 

In the case of V344 Lyr, a strong signal of negative 
superhump was visible from the start of observations (at BJD 2455003) but 
it was weakened before the first superoutburst. 
However a weak signal continued to the next supercycle (SC No. 2) 
but it disappeared around BJD 2455140 before the next superoutburst. 
After a long interval of more than 500~d a new signal of the negative superhump
 again appeared in the early phase of SC No. 7 around BJD 2455670 and it 
stayed almost near the next superoutburst but it disappeared before it. 
  However a weak signal of negative superhump was still visible in the early 
 part of SC No. 8 but it finally disappeared in its later phase. 
 From these observations it will be difficult to tell what initiates 
the disk tilt and what quenches it finally. 

As already noticed and discussed in details by \citet{woo11v344lyr}, a sudden 
excitation of oscillation with a frequency range of negative superhump 
occurred in the descending branch of several normal outbursts in V344 Lyr 
but the oscillation did not grow to a full negative superhump
[see also figure 27 of \citet{woo11v344lyr}].

Here we discuss this kind of events by calling it as an impulsive (failed) 
negative superhump to distinguish it from the ``ordinary'' negative superhump 
discussed in the main part of this paper. The ordinary negative superhump 
has been quite stable (sometimes continued over 300~d) while the impulsive 
negative superhump was short-lived (less than several days) and it never 
developed to a full-scale negative superhump (so we added an 
adjective ``failed'').  \citet{woo11v344lyr} suggested 
from this type of events that the disk tilt could be generated 
by the impulsive coupling between an intensified disk magnetic field 
and the field of the primary or secondary star. 

Here we suggest another possibility for the impulsive negative superhump. 
The impulsive negative superhump is seen in the power spectrum of V344 Lyr 
as a signal which appears near the frequency of negative superhump but 
it quickly moves to higher frequency and disappears. It occurs 
in the descending branch of a normal outburst [figure \ref{fig:v344failnsh};
figure 16 in \citet{woo11v344lyr}].  This quick variation of the
frequencies can be illustrated as follows: in the normal outbursts
shown in figure \ref{fig:v344failnsh}, the negative superhumps
appeared as signals with a longer period, and then
switched to ones with a shorter period [0.0839(1)~d] in outburst No. 2.
[Here the outburst number is that used by \citet{woo11v344lyr} and by 
\citet{can12v344lyr} and we also adopt this number in this paper.]
In outburst No. 19, the signal started with a period
[0.0879(2)~d] close to the orbital period, and then
the period smoothly decreased to 0.0794(2)~d.  The latter period
is astonishingly short, corresponding to $\epsilon^*$=11\%.
There is some indication that these impulsive negative superhump
in outburst No. 2 also started with a period close the orbital period.

As already noted by \citet{woo11v344lyr}, this kind of 
event occurred during outbursts Nos. 2, 10, 17, and 19.
 From figure \ref{fig:v344spec2dlasso},
we add to this list those outbursts of Nos. 28, 42, 54, and 62, and 
possibly Nos. 29, 30, 41, 43, 63.  We also find the impulsive negative 
superhump in three normal outbursts just prior to a superoutburst 
which occurred at around BJD 2456193 in recently 
released Kepler data of V344 Lyr shown in the last panel of 
figure \ref{fig:v344spec2dlasso}. 
In V1504 Cyg, we found the similar impulsive negative superhump
in a ``failed superoutburst'', which occurred two outbursts
before a superoutburst (figure 77 in \cite{Pdot3}).
We could not find strong evidence for the impulsive negative superhump
in other normal outbursts, but instead we find some sign of 
the failed positive superhump in normal outbursts just prior to 
a superoutburst as already discussed in Paper I. 

It would be worth noting that negative superhumps were reported
twice in normal outbursts of V1159 Ori \citep{pat95v1159ori}.
These negative superhumps had period deficit ($-$6.0--7.9\% in
$\epsilon^*$) unusually large for this orbital period (0.0624~d).
These values were so unusual (see table 2 of \cite{mon09diskprecession}
or table 3 of \cite{ole09j2100}), not all the literature lists
this detection as negative superhumps.  This unusual period deficit
may be understood if observed features were the impulsive negative superhump.

As seen in the global light curve of V344 Lyr 
(see figure \ref{fig:v344spec2dlasso}),  
we find that all these normal outbursts accompanied with the 
impulsive negative superhump listed above occurred just prior to 
the next superoutburst and they look like a kind of lead for 
an impending superoutburst. 
There is no such event in the early phase of a supercycle. 
In fact, all seven superoutbursts observed in V344 Lyr 
we have studied were preceded by one, two or three normal outbursts 
accompanied with the impulsive negative superhump. 
The outbursts with possible impulsive negative superhump in V1159 Ori
also occurred in the later stage of the supercycles,
though not immediately before the next superoutbursts.
Although it is still speculative, we suggest that the impulsive negative 
superhump observed in V344 Lyr may be excited by the 3:1 resonance 
for the tidal inclination (tilt) instability discussed by 
\citet{lub92tilt}.
As shown by him, the condition for the resonant excitation of the tidal 
inclination (tilt) instability at 3:1 resonance is very similar to that for 
the ordinary 3:1 resonance tidal eccentric instability but the tilt 
instability occurs in a slightly smaller disk radius than that of the 
ordinary eccentric instability, that means that the tidal tilt instability 
occurs slightly earlier than the tidal eccentric instability. 
This fits very well with the picture presented above. 

However there remains a problem for this explanation,  as the growth rate 
for the tilt instability is much low as compared with the tidal eccentric 
instability as pointed out by \citet{lub92tilt} and it is not at all clear 
if the tilt of disk could be produced within a time-scale of the outburst 
duration. Here we leave it as a future problem to be solved. 
As for a question whether or not the impulsive negative 
superhump could be a mechanism to produce the ordinary negative superhump 
if some other condition could be met, we do not know its answer. 
As a matter of fact, the impulsive negative superhump, that occurred 
during outbursts No. 2 and No. 62, rather worked for quenching 
the existing negative superhump. 

\section{Summary}
We have analyzed the short-cadence Kepler light curves of two SU UMa stars, 
V344 Lyr and V1504 Cyg extending for an interval of well over two years 
(from BJD 2455003--2455832) to study their superoutbursts and superhumps. 
The followings are our major findings.

(1) We have made two-dimensional power spectral analysis for the Kepler 
light curves of V344 Lyr as well as V1504 Cyg. The light curve of V344 Lyr 
and its 
power spectra show both similarity and difference with those of V1504 Cyg 
studied in Paper I. The negative superhump signals seen well in V1504 Cyg 
were found in V 344 Lyr as well but they were more patchy 
 (i.e., come and go)  as compared with those of V1504 Cyg. 
 The correlation between an appearance of the negative superhump 
 and the reduction of outburst frequency, which is well established 
 in the case of V1504 Cyg, 
 exists in the case of V344 Lyr but it is much weaker.
 As for V1504 Cyg, additional Kepler data basically confirm the findings of 
 Paper I. 
 
 (2)  We have presented detailed analysis of frequency variation of 
 the negative superhump for a complete supercycle No. 4 of V1504 Cyg 
 because \citet{sma13negSH} criticized our results in Paper I. We have shown 
 by using the $O-C$ diagram that a frequency jump of the negative superhump 
 takes place during a rising branch of an outburst 
 with a short time scale most likely less than 0.5 d
 exactly as predicted by the disk instability model. We also find that 
 amplitudes of the negative superhump vary systematically during 
 a supercycle in such a sense that its amplitude follows the same pattern with 
 the light curve with a smaller scale. We suggest that these variations are 
 caused by an addition of the disk component during outbursts 
 to the superhump light source besides that of gas stream component 
 as evidenced by the phase-average light profiles of 
 the negative superhump during quiescence, superoutburst, and normal outburst. 
 
 (3) We have presented a new two-dimensional power spectral analysis based on 
  a new method called ``Lasso'' for individual supercycles of 
  V1504 Cyg and V344 Lyr. The new method gives very sharp peaks in the power 
  spectra and thus it is very suitable for the study of frequency variation 
  in cataclysmic variable stars. 
  
  (4) We have analysed frequency variations of the positive and negative 
  superhumps, in particular simultaneous variations of these two superhumps. 
  If they are appropriately converted, it is found that they vary in unison. 
  This suggests strongly that these variations represent disk radius variation 
  during the supercycle of SU UMa stars, supporting the thermal-tidal 
  instability model for the superoutburst and supercycle of SU UMa stars. 
  
  (5) We have examined frequency (or period) variations of the positive 
  superhumps during superoutbursts in V1504 Cyg and V344 Lyr. 
  Two different factors contribute to the frequency variations of 
  the positive superhumps: the disk radius variation and the pressure effects. 
  As for the pressure effects, the higher 
  the disk temperature the slower the precession rate, because the 
  eccentricity wave propagates in a wider range of disk radii. Observations 
  of the positive superhump during a superoutburst of V344 Lyr 
  exhibit a characteristic variation in that its period (or its precession 
  rate) decreases rapidly 
  from its highest value in its start and then it decreases more slowly 
  during the plateau stage of superoutburst, reaching a local 
  minimum and it again increases to a local maximum near 
  the end of the superoutburst when the nature of the superhump changes 
  from the disk origin to the gas stream origin (i.e., the late superhump).
  By using the frequency variation of the negative superhump, we can 
  separate the pressure effects on the frequency (or period) variation of the 
  positive superhump during superoutbursts 
  from the effects of disk radius variation.
  
  (7) As already noted by \citet{woo11v344lyr}, a sudden excitation of 
  oscillation with 
  a frequency range near the negative superhump occurred in the descending 
  branch of several outbursts in V344 Lyr but it damped within a few days. 
  This phenomenon was observed exclusively in V344 Lyr and we called it 
  as an impulsive (failed) negative superhump to distinguish from the ordinary 
  negative superhump. These events seem to occur in outbursts just prior to 
  the next superoutburst. We speculate that the impulsive negative superhump 
  might be excited by the 3:1 resonance for the tidal inclination (tilt) 
  instability first discussed by \citet{lub92tilt} and it seems to act 
  as a kind of ``lead'' of the impending superoutburst.    

\medskip

Note added in proof (2013 September 10):
We have replaced table 2 of the original version to a new one because there
were some errors. In particular, the number of normal outbursts in SC No. 3
of V1504 Cyg is 9 instead of 10. In fact, this error already occurred
in table 2 of Paper I (Osaki and Kato, PASJ, 65, 50, 2013). However,
this correction does not affect any discussions in Paper I.

\medskip

This work was supported by the Grant-in-Aid
``Initiative for High-Dimensional Data-Driven Science through Deepening
of Sparse Modeling'' from the Ministry of Education, Culture, Sports, 
Science and Technology (MEXT) of Japan.
We thank the Kepler Mission team and the data calibration engineers for
making Kepler data available to the public. We are grateful for the referee, 
Dr. M. Montgomery, for her constructive comments to our paper. 

\appendix

\section{Nodal Precession Rate of a Tilted Disk for Different Mass 
Distribution in the Disk}

It is known that the nodal precession rate of a tilted disk is determined 
by the mass distribution within the disk and its general expression is  
written as (see \cite{lar98XBprecession})
\begin{equation}
\omega_{\rm nPR}=2\pi \nu_{\rm nPR}=
-\frac{3}{4}\frac{GM_2}{A^3} 
\frac{\int\Sigma r^3 dr}{\int\Sigma \Omega r^3 dr} \cos \theta, 
\label{equ:nopr_general}
\end{equation}
where $\omega_{\rm nPR}=2\pi \nu_{\rm nPR}$ is angular frequency of nodal 
precession, $A$ is the binary separation, $\Omega=\sqrt{GM_1/r^3}$ is 
the Keplerian angular rotation rate of the disk, and integrations should be 
performed from the inner edge $r_i$ to the outer edge $R_d$ of the disk. 
If we adopt a simple power-law form with $\Sigma(r)=\Sigma_0 (r/R_d)^n$ 
for the mass distribution within the disk where $n$ is a power law index, 
we can perform integrations in equation (\ref{equ:nopr_general}), 
to obtain 

\begin{equation}
\frac{\nu_{\rm nPR}}{\nu_{\rm orb}}=-\frac{3}{4}\frac{2.5+n}{4+n}\frac{q}
{\sqrt{1+q}}(R_d/A)^{3/2}. 
\label{equ:nopr_n}
\end{equation}
where we assumed $R_d \gg r_i$ and $n>-1.5$.

If we denote the coefficient appearing in equation (\ref{equ:nopr_n}) by 
$c_1=3(2.5+n)/4(4+n)$, we can now discuss it for several 
different mass distribution of the disk. For instance, if we assume 
a constant surface density $\Sigma={\rm const}$, we obtain 
15/32 for the coefficient of equation (\ref{equ:nopr_n}) and we recover 
Montgomery's expression. In the same way, if we assume $n=-1/2$, we recover 
the original expression of equation (\ref{equ:noprecession}).  As discussed 
by \citet{osa89suuma}, the surface density distribution of the disk 
in quiescence in its start (i.e., at the end of outbursts which applies 
for a normal outburst as well as for a superoutburst) is approximately 
given by $\Sigma\propto r$ and we then find  $c_1=21/40$. 
In the same way for a hot quasi-steady disk 
corresponding to the plateau stage of a superoutburst of SU UMa stars, 
the surface density distribution is approximately given by 
$\Sigma\propto r^{-3/4}$ and we find $c_1=21/52$. 
We summarize these results for a correction factor $\eta$ 
in table \ref{tab:eta}. 
The correction factor $\eta$ for different mass distributions is rather 
small, typically less than $10\%$. 

\begin{table}
\caption{A correction factor $\eta$ for different mass distribution in the 
disk.}
\label{tab:eta}
\begin{center}
\begin{tabular}{cccccccccc}
\hline
$n$  &  $-$3/4  & $-$1/2 &  0 & 1  \\
\hline
$c_1$ &  21/52 & 3/7 & 15/32 & 21/40 \\
$\eta$  &  0.94 & 1 & 1.09 & 1.22 \\
\hline
\end{tabular}
\end{center}
\end{table}

On the other hand, since the correction factor $\eta$ 
in the case of quiescence at its start amounts for about 20\%,
some more discussion is needed. Although we know fairly well 
mass distribution of the disk in quiescence at its start, 
we do not know how mass distribution changes within the disk 
during quiescence. First of all, we do not know what proportion of the gas 
stream arrives at the disk rim and what proportion spills over the rim and 
arrives at the inner part in a tilted disk. Furthermore we do not know 
how viscous diffusion changes the mass distribution within the disk during 
quiescence. Nevertheless, we have some restriction for the mass distribution 
within the disk during quiescence. It is well known in the disk instability 
model as represented by an S-shaped thermal equilibrium curve, 
there are two critical surface densities, 
$\Sigma_{\rm min}$ and $\Sigma_{\rm max}$, where $\Sigma_{\rm min}$ 
is that below which no hot state exists, and  $\Sigma_{\rm max}$ is that above 
which no cold state exists. The radius dependence of these two critical ones 
is approximately  linear, i.e., 
$\Sigma_{\rm min}(r)$, $\Sigma_{\rm max}(r) \propto r$. 
As shown by \citet{osa89suuma}, the surface 
density distribution in quiescence at its start $\Sigma_{\rm qstart}$ 
is approximately given by $\Sigma_{\rm qstart} 
\simeq 2\Sigma_{\rm min}$. Then the density distribution begins to 
change by mass addition from the gas stream and by viscous diffusion. 
On the other hand the local surface density $\Sigma (r)$, which starts from 
$2\Sigma_{\rm min}$ must be limited by a condition 
$ \Sigma(r) < \Sigma_{\rm max}(r)$ during quiescence. 
Thus we expect that surface density, $\Sigma (r)$ will most likely be limited 
in a range between $\Sigma_{\rm min}$ 
and $\Sigma_{\rm max}$.   In such a case we do not expect any 
extremely different mass distribution during quiescence 
from that of its start. 
 From these considerations we assume for the moment $\eta \sim 1.22$ 
during quiescence. 
  
In paper I we converted the variation in the frequency of nSH for V1504 Cyg 
to variation in the disk radius by using equation (\ref{equ:noprecession}) 
and by assuming a binary mass ratio $q=0.2$. 
Since a proportional constant in the precession rate of a tilted disk 
appears in equation (\ref{equ:noprecession_mod}) by a combination with 
$\eta q/\sqrt{1+q}$, all discussion remains still valid 
even when a correction factor $\eta=1.22$ is applied if we assume the 
binary mass ratio $q\simeq 0.16$ instead of $q=0.2$. A new value for binary 
mass ratio $q\simeq 0.16$ is more appropriate for the observed orbital 
period of V1504 Cyg. 

An introduction of correction factor $\eta=1.22$ in 
comparison between the positive and negative SH frequencies  
in figure \ref{fig:v344posnegper} makes an agreement poor by 
decreasing the data of negative SH by about 20\%. On the other hand, if the 
same correction factor is applied to figure \ref{fig:v1504posnegper3}
for the SC No. 6 of V1504 Cyg, a better agreement is obtained between these 
two $\epsilon^*$ values. 
It is not clear at present moment how serious a disagreement in the case 
of the SC No. 7 of V344 Lyr is, and we leave it as 
a problem to be solved in future.

\end{document}